\documentclass[amssymb,nofootinbib,nobibnotes,aps,prd,preprintnumbers]{revtex4}

 \usepackage{here}

\usepackage[dvipdfmx]{graphicx}
\usepackage{amssymb,amsfonts,amsmath,bm,color,multirow,cases,empheq,hyperref}
\usepackage{mathrsfs}

\definecolor{darkgreen}{rgb}{0,0.5,0.3}

\usepackage{enumerate}
\usepackage{ulem}
\usepackage{afterpage}

\hypersetup{%
 setpagesize=false,
 bookmarksnumbered=true,%
 bookmarksopen=true,%
 colorlinks=true,%
 linkcolor=blue,%
 citecolor=red}

\newcommand{\fr}[1]{\frac{1}{#1}}

\newcommand{\ord}[1]{{\mathcal O}(#1)}

\newcommand{\nonum}{\nonumber\\ }

\newcommand{\cout}[1]{}

\newcommand{\arrayL}[1]{\left(\begin{array}{#1}}
\newcommand{\arrayR}{\end{array}\right)}
\newcommand{\arrayLb}[1]{\left[\begin{array}{#1}}
\newcommand{\arrayRb}{\end{array}\right]}

\newcommand{\tbeta}{{\tilde{\beta}}}

\begin{document}

\title{New construction of a vacuum doubly rotating black ring \\by the Ehlers transformation} 

\author{Ryotaku Suzuki}
\email{sryotaku@toyota-ti.ac.jp}
\author{Shinya Tomizawa}
\email{tomizawa@toyota-ti.ac.jp}
\affiliation{\vspace{3mm}Mathematical Physics Laboratory, Toyota Technological Institute\vspace{2mm}\\Hisakata 2-12-1, Tempaku-ku, Nagoya, Japan 468-8511\vspace{3mm}}

\begin{abstract}

Using the Ehlers transformation, we derive an exact solution for a doubly rotating black ring in five-dimensional vacuum Einstein theory. 
It is well-known that the vacuum Einstein theory with three commuting Killing vector fields can be reduced to a non-linear sigma model with $SL(3,{\mathbb R})$ target space symmetry.
As shown previously by Giusto and Saxena, the $SO(2,1)$ subgroup in the $SL(3,{\mathbb R})$ can generate a rotating solution from a static solution while preserving asymptotic flatness. This so-called Ehlers transformation actually transforms the five-dimensional Schwarzschild black hole into the five-dimensional Myers-Perry black hole.
However, unlike the case with the black hole, applying this method directly to the static black ring or the Emparan-Reall black ring, does not yield a regular rotating black ring due to the emergence of a Dirac-Misner string singularity.
To solve this undesirable issue, we use a singular vacuum solution of a rotating black ring/lens that already possesses a Dirac-Misner string singularity as the seed solution for the Ehlers transformation. 
The resulting solution is regular, indicating the absence of curvature singularities, conical singularities, orbifold singularities, Dirac-Misner string singularities, and closed timelike curves both on and outside the horizon. 
We show that this solution obtained by the Ehlers transformation coincides precisely with the Pomeransky-Sen'kov solution.
We expect that applying this method to other theories may lead to the finding of new exact solutions, such as solutions for black lenses and capped black holes, as well as  black ring configurations.

\end{abstract}

\date{\today}
\preprint{TTI-MATHPHYS-30}

\maketitle

\section{Introduction}

 The study of black hole solutions in Einstein gravity has proven to be an invaluable platform for examining both classical and quantum gravitational theories. 
 Specifically, the exploration of higher-dimensional black holes has captured significant scientific interest over the past twenty years. 
 This attention is highlighted by achievements such as the microscopic derivation of Bekenstein-Hawking entropy~\cite{Strominger:1996sh} and the theoretical possibility of creating black holes in accelerators due to the presence of large extra dimensions~\cite{Argyres:1998qn}.
Despite significant progress in understanding higher-dimensional black holes, our comprehension of higher-dimensional gravity remains superficial, as it encompasses a broader spectrum of complexities and degrees of freedom. 
According to the topology theorem for stationary black holes in five dimensions~\cite{Galloway:2005mf,Cai:2001su,Hollands:2007aj}, the topology of the event horizon's cross-section is restricted to a sphere $S^3$, a ring $S^1 \times S^2$, or lens spaces $L(p,q)$, provided the spacetime is asymptotically flat and supports two commuting axial Killing vector fields. 
 In the first two cases, we possess established exact solutions to the vacuum Einstein equations~\cite{Tangherlini:1963bw, Myers:1986un, Emparan:2001wn, Pomeransky:2006bd}. 
 By contrast, securing a vacuum black hole solution with the lens space topology, denoted as 
$L(p,q)$, remains challenging and yet unresolved. 
Various types of black hole solutions theory have so far been found, 
with the help of, in part, recent development of solution generating 
techniques.
However, the classification of those black hole solutions has not yet been achieved, even in five-dimensional vacuum Einstein theory, since the development of the solution-generating methods is still insufficient.

 \medskip
 Emparan and Reall~\cite{Emparan:2001wn} first found the exact solution for an $S^1$-rotating black ring, demonstrating that five-dimensional vacuum Einstein theory admits the existence of an $S^1$-rotating spherical black hole and two rotating black rings with identical conserved charges, thereby clearly showcasing the non-uniqueness feature in higher dimensions. In general, since in five dimensions, black holes can rotate around two independent axes, it is natural to consider whether there exists a more general black ring solution with two independent rotations, $S^1$ rotation and $S^2$ rotation.
The first derivation of a black ring with $S^2$-rotation was independently achieved by Mishima and Iguchi~\cite{Mishima:2005id} and Figueras~\cite{Figueras:2005zp} by different methods, though their solutions, which are the same, contain conical singularities. 
The inverse scattering method (ISM), initially developed by Belinski and Zakharov~\cite{Belinsky:1979mh}, has proven indispensable for constructing the most general black ring solutions that rotate along both the $S^1$ and $S^2$ axes.
However, direct application of the ISM to higher dimensions tends to yield singular solutions. Pomeransky adeptly modified the method, successfully applying it to five-dimensional black holes by deriving the five-dimensional Myers-Perry solution from the five-dimensional Schwarzschild solution~\cite{Pomeransky:2005sj}. The ISM was also shown to facilitate the generation of $S^2$-rotating black rings~\cite{Tomizawa:2005wv}. Nevertheless, crafting the $S^1$-rotating black ring using the ISM has proven significantly more challenging due to the inevitable emergence of naked curvature singularities from a regular seed solution.
A significant advancement in this direction was made in Refs.~\cite{Iguchi:2006rd,Tomizawa:2006vp}, which found the appropriate seed, a certain singular solution, to generate the $S^1$-rotating black ring. This progress enabled the creation of the doubly rotating black ring using the ISM, and consequently, Pomeransky and Sen'kov successfully obtained the balanced doubly rotating black ring solution~\cite{Pomeransky:2006bd}. Though they gave only the balanced solution in their paper~\cite{Pomeransky:2006bd}, the unbalanced generalization of the black ring was later presented explicitly in Ref.~\cite{Morisawa:2007di}. Furthermore, a more compact form of this solution was presented by Ref.~\cite{Chen:2011jb}.

\medskip

Meanwhile, the sigma model approach has been established as one of the most effective methods for generating exact solutions to Einstein's equations~\cite{exact}. 
In vacuum Einstein theory with a single Killing vector, this theory can be reduced to a non-linear sigma model coupled with three-dimensional gravity through dimensional reduction~\cite{Ernst:1967wx}. 
Here, the target space of this sigma model is represented by the coset $SL(2, \mathbb{R})/SO(2)$, which facilitates the generation of new solutions by applying group transformations to the coset representative of a seed solution. 
Similarly, for the four-dimensional Einstein-Maxwell theory with a single Killing vector, the corresponding coset is $SU(2,1)/S[U(2) \times U(1)]$~\cite{Ernst:1967by}.
The $SU(2,1)$ symmetry provides two significant transformations: the Ehlers transformation~\cite{Ehlers1957}, which adds a nut charge, and the Harrison transformation~\cite{Harrison1968}, which adds an electric or magnetic charge to a given seed solution. However, the $SU(2,1)$ symmetry cannot simply add rotation to a non-rotating solution while preserving asymptotic flatness. For a stationary and axisymmetric solution, this can be achieved by combining transformations associated with an arbitrarily chosen direction in the two Killing vector space with transitions to other linear combinations of the two Killing vectors. For example, as pointed out by Clement~\cite{Clement:1997tx,Clement:1999bv}, the Kerr solution can be obtained from the Schwarzschild solution using this transformation.
As Maison showed~\cite{Maison:1979kx}, in $d$-dimensional ($d > 4$) vacuum Einstein theory with $(d-3)$ commuting Killing vectors, the coset is represented by $SL(d-2,{\mathbb R})/SO(2,d-4)$. Specifically, for $d=5$, Giusto and Saxena~\cite{Giusto:2007fx} demonstrated that by using the $SO(2,1)$ subgroup of the $SL(3,{\mathbb R})$ symmetry, which preserves asymptotic flatness, one can transform a static solution into a one-parameter family of stationary solutions with angular momentum. 
They illustrated this by deriving the Myers-Perry black hole from the five-dimensional Schwarzschild solution.

\medskip
Additionally,  for five-dimensional minimal supergravity, which is the Einstein-Maxwell-Chern-Simons theory with a specific coupling constant, two Killing vectors with the coset is given by $G_{2(2)}/[SL(2,{\mathbb R}) \times SL(2,{\mathbb R})]$ or $G_{2(2)}/SO(4)$, depending on whether the Killing vectors chosen are a timelike one and a spacelike one, or two spacelike ones, respectively~\cite{Mizoguchi:1998wv,Mizoguchi:1999fu}. Utilizing the $G_{2(2)}$ symmetry, Bouchareb et al.~\cite{Bouchareb:2007ax} developed a method for generating solutions, including an electric Harrison transformation that transforms a five-dimensional vacuum solution into an electrically charged one in five-dimensional minimal supergravity.
Applying this transformation to the five-dimensional vacuum rotating black hole (the Myers-Perry solution~\cite{Myers:1986un}) results in the five-dimensional charged rotating black hole (the Cveti\v{c}-Youm solution~\cite{Cvetic:1996xz}). However, as noted by the authors, applying this Harrison transformation to the vacuum doubly rotating black ring (the Pomeransky-Sen'kov solution~\cite{Pomeransky:2006bd}) does not produce a regular charged doubly spinning black ring solution, as the resulting solution suffers from a Dirac-Misner string singularity.
In our previous work~\cite{Suzuki:2024coe}, we addressed this serious problem and constructed the exact solution for a charged rotating black ring. The key idea was to use the electric Harrison transformation on a vacuum solution of a rotating black ring that possesses a Dirac-Misner string singularity. 
This transformation also provided the first example of a non-BPS exact solution representing an asymptotically flat, stationary spherical black hole, refereed to as a capped black hole, with a domain of outer communication (DOC) having a nontrivial topology, in the sense that the DOC on a timeslice has the topology of $[R^4 \# CP^2] \setminus B^4$~\cite{Suzuki:2023nqf,Suzuki:2024phv}.

 \medskip

 The aim of this paper is to present a new construction of a vacuum solution for a doubly rotating black ring in five-dimensional Einstein theory using the $SL(3,{\mathbb R})$ transformation in a non-linear sigma model approach. 
 The basic procedure resembles the construction of the charged dipole black ring or the capped black hole solution in our previous works~\cite{Suzuki:2023nqf,Suzuki:2024phv}, in which we used the black lens solution which possesses a Dirac-Misner string singularity as a seed solution for the Harrison transformation.
We also use the same seed solution as for the the Harrison transformation---the singular solution with a Dirac-Misner string singularity---for the $SL(3,{\mathbb R})$ transformation.
This singularity will be removed by appropriately adjusting the solution's parameters not before but after the $SL(3,{\mathbb R})$ transformation.
By applying the $SL(3,{\mathbb R})$ transformation to this vacuum solution, we add rotation to the vacuum seed solution. 
After applying the transformation to the seed, we select solution's parameters  appropriately to ensure that the resulting solution is regular, free from Dirac-Misner string singularity, and devoid of conical singularities within the black ring. Consequently, the final solution represents a doubly rotating black ring characterized by its regularity—it does not have curvature singularities, conical singularities, Dirac-Misner string singularities, and orbifold singularities on the rods (which include the rotational axes and the horizon).
According to the uniqueness theorem for vacuum black ring solutions~\cite{Hollands:2007aj,Morisawa:2007di,Tomizawa:2009tb}, our derived solution is expected to be isometric to the well-known Pomeransky-Sen'kov solution~\cite{Pomeransky:2006bd}, provided that the spacetime described by our solution is regular and free from closed timelike curves on and outside the event horizon. However, demonstrating the equivalence of the two solutions is challenging due to the apparent differences arising from the complex and lengthy expressions of our solution, which stem from significant differences in the parameters used in two different solution-generating methods. By analyzing the rod data and asymptotic charges of both solutions, we identify the relationship between the two sets of parameters. Consequently, after correlating the differing parameter sets of the two solutions, we establish that our solution precisely coincides with the Pomeransky-Sen'kov solution.

\medskip

The remainder of this paper is dedicated to constructing the aforementioned black ring solution. In the next section (Sec.~\ref{sec:setup}), we will briefly review the solution-generation method that employs the $SL(3, \mathbb{R})$ invariance to add angular momentum to a vacuum seed solution within the non-linear sigma model framework of five-dimensional Einstein theory.
In Sec.~\ref{sec:ism}, we will introduce the vacuum solution of the rotating black lens, which includes a Dirac-Misner string singularity. This solution will serve as the seed for the $SL(3, \mathbb{R})$ transformation. Previously, this same vacuum solution was utilized in Refs.~\cite{Suzuki:2023nqf, Suzuki:2024phv} to construct a charged black ring solution and a capped black hole, employing it as a seed for the Harrison transformation. In alignment with these earlier works~\cite{Suzuki:2023nqf, Suzuki:2024phv}, we will retain the Dirac-Misner string singularity, choosing not to eliminate it.
In Sec.~\ref{sec:sol}, we will apply the 
$SL(3,{\mathbb R})$-transformation to the vacuum solution, leading to a corresponding rotating vacuum solution. By setting appropriate boundary conditions on the parameters, we will ensure the absence of any singularities—including curvature, conical, and Dirac-Misner string singularities—on the rotational axes and the horizon. This will allow us to derive a doubly rotating black ring solution in the C-metric form. Initially, this solution appears to differ from the well-known Pomeransky-Sen'kov solution. 
However, in Sec.\ref{sec:PS0}, we will show that the resulting solution exactly aligns with the Pomeransky-Sen'kov solution. Finally, in Sec.~\ref{sec:sum}, we will summarize our findings and outline future research directions

\section{Preliminary}\label{sec:setup}
Under the assumption of space-time symmetry, the Einstein equations simplify to equations for scalar fields coupled with a three-dimensional gravity. 
We then review that this system of scalar fields is governed by a nonlinear sigma model.
\subsection{Vacuum Einstein gravity with two commuting Killing vectors}
First, we consider a spacetime  with two mutually commuting Killing vector fields $\xi_a \ (a=0,1)$ such that $[\xi_a,\xi_b]=0$.
Then, by introducing the coordinates $x^a$ as $\xi_a = \partial/\partial x^a$, the metric  can be written as     
\begin{eqnarray}
ds^2 = \lambda_{ab}(dx^a+a^a{}_idx^i)(dx^b+a^b{}_jdx^j) 
      +\tau^{-1}h_{ij}dx^idx^j \,, 
\end{eqnarray}
where  the functions, $\lambda_{ab}=(\xi_a|\xi_b)$, $\tau:=-{\rm det}(\lambda_{ab})$, $a^a{_i}$, the three-dimensional metric $h_{ij}$ ($i=2,3,4$) are independent of the coordinates~$x^a$.  
Let us define the twist one-forms by 
\begin{equation} 
V_a=*(\xi_1\wedge \xi_2\wedge d\xi_a) \,.,
\end{equation}
whose exterior derivative can be written as 
\begin{eqnarray}
dV_a &=& 2*(\xi_1\wedge \xi_2\wedge R(\xi_a)),\label{eq:dV}
\end{eqnarray}
where $R(\xi_a)$ is the Ricci one-form. 
Therefore, by the vacuum Einstein equation, since the right-hand side in Eq.~(\ref{eq:dV}) vanishes, there locally exists the twist potentials $\omega_a$ such that 
\begin{eqnarray}
 d\omega_a=V_a,
 \label{eq:twistpotential} 
\end{eqnarray} 
which can be written as
\begin{eqnarray}
 \partial_k\omega_{a}=\tau \sqrt{|h|} \lambda_{ab} \varepsilon_{ijk}  h^{im}h^{jn}\partial_m a^b{}_n.\label{eq:domega}
\end{eqnarray}
Then, the vacuum Einstein equations can be expressed as
the field equations for the five scalar fields, $\{\lambda_{ab},\omega_a\}$, 
\begin{eqnarray}
\Delta_h \lambda_{ab}&=&\lambda^{cd} h^{ij}\frac{\partial \lambda_{ac}}{\partial x^i} \frac{\partial \lambda_{bd}}{\partial x^j}+\tau^{-1} h^{ij}\frac{\partial \omega_a}{\partial x^i} \frac{\partial \omega_b}{\partial x^j},\label{eq:eom1}\\
\Delta_h \omega_{a}&=& \tau^{-1} h^{ij}\frac{\partial \tau}{\partial x^i} \frac{\partial \omega_a}{\partial x^j}+\lambda^{bc} h^{ij}\frac{\partial \lambda_{ab}}{\partial x^i} \frac{\partial \omega_c}{\partial x^j},\label{eq:eom2}
\end{eqnarray}
and the Einstein equations for the three-dimensional metric $h_{ij}$, which is coupled with the five scalar fields,
\begin{eqnarray}
R^h_{ij} &=&  \frac{1}{4} \lambda^{ab}\lambda^{cd}
              \frac{\partial \lambda_{ac}}{\partial x^i }  \frac{\partial \lambda_{bd}}{\partial x^j } 
   + \frac{1}{4}\tau^{-2}\frac{\partial \tau}{\partial x^i} \frac{\partial \tau}{\partial x^j }   
    -\frac{1}{2}\tau^{-1}\lambda^{ab} \frac{\partial \omega_a}{\partial x^i }\frac{\partial \omega_b}{\partial x^j },       
\label{eq:Rij}
\end{eqnarray} 
where $\Delta_h$ is the Laplacian and $R^h_{ij}$ denotes the Ricci tensor with respect to $h_{ij}$, respectively.

\medskip
\subsubsection{$SL(3,{\mathbb R})$ nonlinear sigma model}
Thus, as a consequence of the existence of two isometries $\xi_a$, we have five scalar fields $\lambda_{ab},\omega_a$ 
$(a=0,1)$, which we denote collectively by coordinates $\Phi^A=(\lambda_{ab},\omega_a)$. 
Then, we can find that the equations of motion, eqs.~(\ref{eq:eom1}), (\ref{eq:eom2}) and (\ref{eq:Rij})
are derived from the following action for sigma-model $\Phi^A$ coupled with three-dimensional gravity with respect to the metric $h_{ij}$, 
\begin{eqnarray}
S=\int\left(R^h
 -G_{AB}\frac{\partial \Phi^A}{\partial x^i}
 \frac{\partial \Phi^B}{\partial x^j}h^{ij}\right)\sqrt{|h|}d^3x \,,\label{eq:action}
\end{eqnarray}
where $R^h$ denotes the Ricci scalar with respect to $h_{ij}$, and the target space metric, $G_{AB}$, is given by 
\begin{eqnarray}
G_{AB}d\Phi^Ad\Phi^B 
&=& \frac{1}{4}{\rm Tr}(\lambda^{-1}d\lambda\lambda^{-1}d\lambda )
   + \frac{1}{4}\tau^{-2}d\tau^2 
    -\frac{1}{2}\tau^{-1}V^T\lambda^{-1}V,
\end{eqnarray}
with $\lambda=(\lambda_{ab})$, $\omega=(\omega_1,\omega_2)^T$ and $V=d\omega$.
It can be actually shown that varying the action by $h_{ij}$ derives the equation~(\ref{eq:Rij}),
\begin{eqnarray}
R^h_{ij} &=& G_{AB}\frac{\partial \Phi^A}{\partial x^i} 
                 \frac{\partial \Phi^B}{\partial x^j},  
      \end{eqnarray} 
and also varying the action by $\Phi^A$ can derive the equations~(\ref{eq:eom1}) and (\ref{eq:eom2}),
\begin{eqnarray}
\Delta_h\Phi^A+h^{ij}\Gamma^A_{BC}\frac{\partial \Phi^B}{\partial x^i} 
                                  \frac{\partial \Phi^C}{\partial x^j} 
  =0, \label{eq:harmonic_map}
\end{eqnarray}
where  $\Gamma^A_{BC}$ is the Christoffel symbol with respect to $G_{AB}$.

\subsubsection{Coset matrix}
Maison~\cite{Maison:1979kx} showed that the action for the scalar fields $\Phi^A$ is invariant under the global $SL(3,{\mathbb R})$ transformation, introducing the 
$SL(3,{\mathbb R})$ matrix 
$\chi$,  defined by the $3\times3$ matrix
\begin{eqnarray}
\chi= \left(
  \begin{array}{ccc}
  \displaystyle \lambda-\frac{\omega \omega^T}{\tau} &  \displaystyle \frac{\omega}{\tau}\\
   \displaystyle  \frac{\omega ^T}{\tau}                     & \displaystyle  -\frac{1}{\tau}
    \end{array}
 \right) \,,
\end{eqnarray}
where this matrix $\chi$ is symmetric, $\chi^T=\chi$, and unimodular, $\det(\chi)= 1$.
Then, the action (\ref{eq:action}) can be written in terms of $\chi$ as follows
\begin{eqnarray}
S=\int\left(R^h
 - \frac{1}{4}h^{ij} {\rm tr}(\chi^{-1}\partial_i\chi \chi^{-1}\partial_j\chi)
\right)\sqrt{|h|}d^3x \,,
\end{eqnarray}
which  is invariant under the transformation
\begin{eqnarray}
\chi\to \chi'=g \chi g^T,\quad h\to h \label{eq:sl3r}
\end{eqnarray}
with $g\in SL(3,{\mathbb R})$.
The equations of motion~(\ref{eq:Rij}), (\ref{eq:eom1}) and  (\ref{eq:eom2})  can be written, in terms of $\chi$, as
\begin{eqnarray}
&&d\star_3 (\chi^{-1} d\chi)=0,\\
&&R^h_{ij}=\frac{1}{4}{\rm tr}(\chi^{-1}\partial_i\chi \chi^{-1}\partial_j\chi).
\end{eqnarray}
Thus, the presence of two commuting Killing vector
fields reduces the five-dimensional vacuum Einstein theory to a three-dimensional non-linear sigma model with a $SL(3,{\mathbb R})$ target space symmetry. 
If both two Killing vectors are spacelike, it is described by the $SL(3,{\mathbb R})/SO(3)$ sigma model coupled to gravity, while
if one of the two Killing vectors is timelike, the symmetry is replaced with $SL(3,{\mathbb R})/SO(2,1)$~\cite{Maison:1979kx}.

\subsection{Einstein gravity with three commuting Killing vectors}
It was  seen from the previous subsection that assuming the existence of one timelike Killing vector $\xi_0 = \partial/\partial t$ and one spacelike axial Killing vector $\xi_1=\partial/\partial\psi$, this theory reduces to the $SL(3,{\mathbb R})/SO(2,1)$ non-linear sigma models coupled to three-dimensional gravity~\cite{Maison:1979kx}. 
Furthermore, if we assume the presence of another spacelike Killing vector $\xi_2=\partial/\partial x_2$ such that three Killing vectors mutually commute, $[\xi_I,\xi_J]\ (I,J=0,1,2)$, it can be shown that the Einstein gravity can be completely reduced to the non-linear sigma model as follows: 
Under a further assumption of the presence of the third spacelike axial Killing vector $\xi_2=\partial/\partial x^2$, the two-dimensional surface orthogonal to three $\xi_I\ (I,J=0,1,2)$ is integrable, as 
shown in Ref.~\cite{weyl,Harmark}.
Under the symmetry assumptions, the metric can be written in the Weyl-Papapetrou form:
\begin{eqnarray}
ds^2
&=& \lambda_{ab}(dx^a+a^a{}_2dx^2)(dx^b+a^b{}_2dx^2)+\tau^{-1}\rho^2 (dx^2)^2
     +\tau^{-1}e^{2\sigma}(d\rho^2+dz^2) \,, \label{eq:WPform}
\end{eqnarray}
where the coordinates $x^a$ ($a=0,1$) and $x^2$ denote the Killing coordinates and all functions $\lambda_{ab}$, 
$\tau:=-{\rm det}(\lambda_{ab})$, $a^a{}_2$, $\sigma$ are independent of $x^a$ and $x^2$. 
The coordinates $(\rho,z)$ that span a two-dimensional base space, $\Sigma=\{(\rho,z)|\rho\ge 0,\ -\infty<z<\infty \}$, 
are globally well-defined, harmonic, and mutually conjugate on $\Sigma$.

\medskip 
In this coordinate system, the equation of motion~(\ref{eq:harmonic_map}) for $\Phi^A$ reduces to
\begin{eqnarray}
\Delta_\gamma \Phi^A + 
\Gamma^A_{BC}[\Phi^B_{,\rho}\Phi^C_{,\rho}+\Phi^C_{,z}\Phi^C_{,z}]=0 \,, 
\label{eq:scalar}
\end{eqnarray}
where $\Delta_\gamma$ is the Laplacian with respect to the abstract 
three-dimensional metric $\gamma=d\rho^2+dz^2+\rho^2d\varphi^2$. 
On the other hand, the function $\sigma$ is determined by 
\begin{eqnarray}
\frac{2}{\rho}\sigma_{,\rho}&=&R^h_{\rho\rho}-R^h_{zz}\nonumber\\
 &=&G_{AB}[\Phi^A_{,\rho}\Phi^B_{,\rho}-\Phi^A_{,z}\Phi^B_{,z}]\notag\\
&=&\frac{1}{4}\tau^{-2}[(\tau_{,\rho})^2-(\tau_{,z})^2]+\frac{1}{4}\lambda^{ab}\lambda^{cd}(\lambda_{ac,\rho}\lambda_{bd,\rho}-\lambda_{ac,z}\lambda_{bd,z} )-\frac{1}{2}\tau^{-1}\lambda^{ab}(\omega_{a,\rho}\omega_{b,\rho}-\omega_{a,z}\omega_{b,z}),\label{eq:sigmarho}   \\
   \frac{1}{\rho}\sigma_{,z}&=&R^h_{\rho z},\nonumber\\
                            &=&G_{AB}\Phi^A_{,\rho}\Phi^B_{,z}\notag\\
                            &=&\frac{1}{4}\tau^{-2}\tau_{,\rho}\tau_{,z}+\frac{1}{4}\lambda^{ab}\lambda^{cd}\lambda_{ac,\rho}\lambda_{bd,z}-\frac{1}{2}\tau^{-1}\lambda^{ab}\omega_{a,\rho}\omega_{b,z},
\label{eq:sigmaz}                            
\end{eqnarray}
where the integrability of the function $\sigma$, $\sigma_{,\rho z}=\sigma_{,z \rho }$, is assured by Eq.~(\ref{eq:scalar}). 
From Eq.~(\ref{eq:twistpotential}) or (\ref{eq:domega}), the metric functions $a^a{}_{2}\ (a=0,1)$ are determined by 
\begin{align}
\partial_\rho a^a{}_2 =- \frac{\rho }{\tau} \lambda^{ab}\partial_z \omega_{b},\quad \partial_z a^a{}_2= \frac{\rho }{\tau} \lambda^{ab}\partial_\rho \omega_{b}.\label{eq:a0phi-neutral}
\end{align}
Thus, once one determines the five scalar fields $\Phi^A= (\lambda_{ab},\omega_a)$ through Eq.~(\ref{eq:scalar}), one can obtain the solutions to the five-dimensional vacuum Einstein equations with three Killing isometries.  
It turns out that Eq.~(\ref{eq:scalar}) are derived from the following action
\begin{eqnarray}
S&=&\int d\rho dz \rho\left[ G_{AB}(\partial\Phi^A)(\partial\Phi^B)\right] 
\nonumber \\
 &=& \int d\rho dz \rho 
     \biggl[\:  
             \frac{1}{4}{\rm Tr}(\lambda^{-1}\partial\lambda\lambda^{-1}
                                             \partial\lambda )
           + \frac{1}{4}\tau^{-2}\partial\tau^2 
                   - \frac{1}{2}\tau^{-1}v^T\lambda^{-1}v 
                \biggr]  \\
&=&  \frac{1}{4}\int d\rho dz 
    \rho {\rm tr}\left(\chi^{-1}\partial_i\chi\chi^{-1}\partial^i\chi\right)\,. 
\label{eq:action} 
\end{eqnarray} 
In terms of the coset matrix $\chi$, Eqs.~(\ref{eq:scalar}) and (\ref{eq:sigmarho}), (\ref{eq:sigmaz}) can be written, respectively,  as
\begin{eqnarray}
 \partial_\rho( \rho \partial_\rho \chi \chi^{-1}) + \partial_z( \rho \partial_z \chi \chi^{-1}) = 0,
\end{eqnarray}
and
\begin{align}
\frac{2}{\rho}\sigma_{,\rho}&=  - \fr{\rho^2} + \fr{4\rho^2} {\rm tr} [( \rho \partial_\rho \chi \chi^{-1})^2-( \rho \partial_z \chi \chi^{-1}) ^2],\\
\frac{1}{\rho}\sigma_{,z}&=   \fr{4\rho^2} {\rm tr}[( \rho \partial_\rho \chi \chi^{-1})( \rho \partial_z \chi \chi^{-1})] . \label{eq:f}
\end{align}

\subsection{Ehlers transformation}\label{sec:GStr}

When one constructs a new asymptotically flat vacuum solution from a certain vacuum seed by using the $SL(3,\mathbb{R})$ transformation, it is convenient to use a special $SL(3,\mathbb{R})$ transformation  that preserves the asymptotic metric at infinity.  
This ensures that the five-dimensional Minkowski metric, which can be expressed in terms of the angular coordinates $(\psi,\phi)$ with the periodicity $2\pi$,
\begin{align}
ds^2 &= - dt^2 + dr^2 + r^2 (d\theta^2 + \sin^2\theta d\psi^2+\cos^2\theta d\phi^2),,\label{eq:flat}
 \end{align}
 is maintained, or,  alternatively,  in terms of the Euler angles $\phi_\pm:=\psi\pm \phi$ and $\Theta:=2\theta$, it can be also written as
\begin{align}
ds^2        &=  - dt^2 + dr^2 + \frac{r^2}{4} \left[ (d\phi_++\cos\Theta d\phi_-)^2+d\Theta^2+\sin^2\Theta d\phi_-^2 \right],\label{eq:flat_Euler}
\end{align}

Giusto and Saxena showed in Ref.\cite{Giusto:2007fx} that the $SL(3,{\mathbb R})$-transformation preserving asymptotic flatness is given by the one that leaves the limit $r \to \infty$ of the coset matrix $\chi$ corresponding to the Minkowski metric~(\ref{eq:flat_Euler}):
\begin{eqnarray}\label{eq:coset-inf}
 \lim_{r\to\infty}\chi = \arrayLb{ccc} -1&0&0\\0&0&-1\\0&-1&0 \arrayRb, 
\end{eqnarray}
 rather than the one with the orthogonal angles~(\ref{eq:flat}), invariant under the transformation~(\ref{eq:sl3r})\footnote{It should be noted that $\chi$ at $r \to \infty$ in Ref.\cite{Giusto:2007fx} has the opposite signature to the one we used, due to the difference in the signature of the twist potentials~(\ref{eq:twistpotential}).}.
This can be expressed by the subgroup $SO(2,1)$ within the $SL(3,\mathbb{R})$.
Moreover, among the subgroup with three parameters, only one parameter family results in a nontrivial transformation. 
More precisely, choosing the two Killing vectors $(\xi_0, \xi_1) = (\partial / \partial t, \partial / \partial \psi)$ or $(\xi_0, \xi_1) = (\partial / \partial t, \partial / \partial \phi)$ results in only trivial transformations but selecting other Killing vectors $(\xi_0, \xi_1) = (\partial / \partial t, \partial / \partial \phi_+)$ or $(\xi_0, \xi_1) = (\partial / \partial t, \partial / \partial \phi_-)$ yields a physical transformation that adds angular momentum to a static seed solution, which is expressed as
\begin{align}\label{eq:sl3r-trans}
\chi \to \chi' = g_r^T \chi g_r,\quad g_r = \arrayLb{ccc}
 1&0&\tilde{\beta}\\
 -\tilde{\beta}&1&-\frac{\tilde{\beta}^2}{2}\\
 0&0&1
 \arrayRb \in SL(3,\mathbb{R}).
\end{align}
Under this transformation, the set of the scalar fields $\Phi^A = \{\lambda_{ab},\omega_a\}$ is changed  to the set of different  ones $\Phi'^A = \{ \lambda_{ab}',\omega_a'\}$ as follows
\begin{align}\label{eq:scalar-trans}
\begin{split}
&\lambda'_{00} = D^{-1}\left[ \lambda _{00}-2 \tbeta  \lambda _{01}+\tbeta ^2 \left(\lambda _{11}+\epsilon^{ij}\omega_i \lambda_{j0}\right)-\tbeta ^3 \epsilon^{ij} \omega_i \lambda_{j1} 
+\frac{\tbeta ^4 \left(\tau 
   \left(-\omega _0^2+\tau  \lambda _{00}\right)+\left(\epsilon^{ij}\omega_i \lambda_{j0}\right){}^2\right)}{4 \lambda _{00}}\right],\\
&\lambda'_{01}=D^{-1}\left[ \lambda_{01}-\tbeta  \left(\lambda_{11}+\epsilon^{ij}\omega_i \lambda_{j0}\right)+\frac{3}{2} \tbeta ^2 \epsilon^{ij}\omega_i \lambda_{j1}
-\frac{\tbeta ^3 \left(\tau    \left(-\omega _0^2+\tau  \lambda_{00}\right)+\left( \epsilon^{ij} \omega_i \lambda_{j0}\right){}^2\right)}{2 \lambda_{00}} \right],\\
&\lambda'_{11}=D^{-1} \left[\lambda_{11}-2 \tbeta \epsilon^{ij} \omega_i \lambda_{j1} +\frac{\tbeta ^2 \left(\tau  \left(-\omega _0^2+\tau  \lambda
   _{00}\right)+\left(\epsilon^{ij}\omega_i \lambda_{j0}\right){}^2\right)}{\lambda_{00}}\right],\\
&\omega'_0=D^{-1}\left[\omega _0+\tbeta  \left(-\omega _1-\omega _0^2+\tau  \lambda
   _{00}\right)+\frac{3}{2} \tbeta ^2 \left(\omega _0 \omega _1-\tau  \lambda
   _{01}\right)+\frac{1}{2} \tbeta ^3 \left(-\omega _1^2+\tau\lambda _{11} 
   \right)\right],\\
&\omega'_1=D^{-1}\left[\omega _1+\tbeta  \left(-\omega _0 \omega _1+\tau  \lambda
   _{01}\right)+\frac{1}{2} \tbeta ^2 \left(\omega _1^2-\tau\lambda _{11} 
   \right)\right],\\
& \tau' = \tau/D,   
\end{split}
\end{align}
with
\begin{align}
D = 1-2 \tbeta  \omega _0+\tbeta ^2 \left(\omega _1+\omega _0^2-\tau  \lambda
   _{00}\right)-\tbeta ^3 \left(\omega _0 \omega _1-\tau  \lambda
   _{01}\right)+\frac{1}{4} \tbeta ^4 \left(\omega _1^2- \tau\lambda _{11}
   \right),
\end{align}
where the new parameter $\tilde{\beta}$ is physically related to the angular momentum with respect to $\partial/\partial{\phi_+}$ or $\partial/\partial{\phi_-}$, depending on the choice of the Killing vector $\xi_1 = \partial/\partial{\phi_+}$ or $\partial/\partial{\phi_-}$. The one-forms $a^{\prime a}{}_i dx^i$ $(a = 0,1)$ for the transformed solution are determined by the five scalar functions ${\lambda'_{ab}, \omega'_a}$ from Eqs.~(\ref{eq:domega}), after replacing ${\lambda_{ab}, \omega_a}$ with ${\lambda'_{ab}, \omega'_a}$ and substituting Eq.~(\ref{eq:scalar-trans}).
Thus, $a'^a{}_i$ can be determined from $a^a{}_i$ and $\{\lambda_{ab}, \omega_a\}$ by
\begin{align}\label{eq:sol-ai-new0}
\begin{split}
&\partial_m (a'^0{}_n-a^0{}_n)-\partial_n (a'^0{}_m-a^0{}_m) = 
 \sqrt{h} \, \varepsilon_{mn\ell} \, h^{\ell k} \left[ \tbeta ( 2 (\tilde{{\bf X}}^0{}_0)_k + (\tilde{\bf X}^1{}_1)_k ) +\frac{\tbeta^2}{2} (2 (\tilde{\bf Y}^1)_k-(\tilde{\bf X}^0{}_1)_k)+\frac{\tbeta^3}{2} (\tilde{\bf Y}^0)_k\right],\\
&\partial_m (a'^1{}_n-a^1{}_n)-\partial_n (a'^1{}_m-a^1{}_m) =
  \sqrt{h}\, \varepsilon_{mn\ell}\, h^{\ell k}\left[ \tbeta (\tilde{\bf X}^1{}_0)_k + \frac{3}{2} \tbeta^2 (\tilde{\bf X}^0{}_0)_k + \frac{\tbeta^3}{2} ((\tilde{\bf Y}^1)_k-(\tilde{\bf X}^0{}_1)_k)+\frac{\tbeta^4}{4}(\tilde{\bf Y}^0)_k \right],
\end{split}
\end{align}
where the one-forms $(\tilde{\bf X}^a{}_b)_k$ and $(\tilde{\bf Y}^a)_k$ can be expressed in terms of only the scalar fields $\Phi^A$ as : 
\begin{align}\label{eq:sol-xij-yi}
\begin{split}
& \tilde{\bf X}^a{}_b =  \lambda^{ac} d\lambda_{cb} 
 - \tau^{-1} \lambda^{ac} d\omega_c \omega_b + \tau^{-1} \delta^a_1 \delta^0_b  \lambda^{0c} d\omega_c,\\
&  \tilde{\bf Y}^a = 
- \epsilon^{ab} d\omega_b-  \tau^{-1} \epsilon^{ab} \omega_b   \omega_c \lambda^{cd} d\omega_d
 +\epsilon^{ab} d\lambda_{bc} \lambda^{cd} \omega_d+\tau^{-1}\epsilon^{ab} \omega_b d\tau.
 \end{split}
\end{align}
The assumption of a further Killing vector $\xi_2=\partial/\partial \phi_-$ or $\xi_2=\partial/\partial \phi_+$, depending on the choice  of $\xi_1=\partial/\partial \phi_+$ or $\xi_1=\partial/\partial \phi_-$, simplifies  Eqs.~(\ref{eq:sol-ai-new0}) to 
\begin{align}
& \partial_\rho (a'^0{}_2- a^0{}_2)= - \rho \left[ \tbeta ( 2 (\tilde{{\bf X}}^0{}_0)_z + (\tilde{\bf X}^1{}_1)_z ) +\frac{\tbeta^2}{2} (2 (\tilde{\bf Y}^1)_z-(\tilde{\bf X}^0{}_1)_z)+\frac{\tbeta^3}{2} (\tilde{\bf Y}^0)_z\right],\\
& \partial_z (a'^0{}_2- a^0{}_2)= 
\rho  \left[ \tbeta ( 2 (\tilde{{\bf X}}^0{}_0)_\rho + (\tilde{\bf X}^1{}_1)_\rho ) +\frac{\tbeta^2}{2} (2 (\tilde{\bf Y}^1)_\rho-(\tilde{\bf X}^0{}_1)_\rho)+\frac{\tbeta^3}{2} (\tilde{\bf Y}^0)_\rho\right],\\
& \partial_\rho (a'^1{}_2- a^1{}_2)= -\rho \left[ \tbeta (\tilde{\bf X}^1{}_0)_z + \frac{3}{2} \tbeta^2 (\tilde{\bf X}^0{}_0)_z + \frac{\tbeta^3}{2} ((\tilde{\bf Y}^1)_z-(\tilde{\bf X}^0{}_1)_z)+\frac{\tbeta^4}{4}(\tilde{\bf Y}^0)_z \right],\\
& \partial_z (a'^1{}_2- a^1{}_2)=
 \rho \left[ \tbeta (\tilde{\bf X}^1{}_0)_\rho + \frac{3}{2} \tbeta^2 (\tilde{\bf X}^0{}_0)_\rho + \frac{\tbeta^3}{2} ((\tilde{\bf Y}^1)_\rho-(\tilde{\bf X}^0{}_1)_\rho)+\frac{\tbeta^4}{4}(\tilde{\bf Y}^0)_\rho \right],
\end{align}
The integration can be written as
\begin{align}\label{eq:sol-ai-new}
\begin{split}
&  a'^0{_2} = a^0{}_2 + \tbeta ( 2 {\bf X}^0{}_0 + {\bf X}^1{}_1 ) +\frac{\tbeta^2}{2} (2 {\bf Y}^1-{\bf X}^0{}_1)+\frac{\tbeta^3}{2} {\bf Y}^0,\\
& a'^1{}_2 = a^1{}_2 + \tbeta {\bf X}^1{}_0 + \frac{3}{2} \tbeta^2 {\bf X}^0{}_0 + \frac{\tbeta^3}{2} ({\bf Y}^1-{\bf X}^0{}_1)+\frac{\tbeta^4}{4}{\bf Y}^0,
\end{split}
\end{align}
where ${\bf X}^a{}_b$ and ${\bf Y}^a$ are the solutions of
\begin{align}
\begin{split}
&\partial_\rho {\bf X}^a{}_b = - \rho\, (\tilde{\bf X}^a{}_b)_z,\quad \partial_z {\bf X}^a{}_b =  \rho\, (\tilde{\bf X}^a{}_b)_\rho,\\
&\partial_\rho {\bf Y}^a = - \rho\, (\tilde{\bf Y}^a)_z,\quad \partial_z {\bf Y}^a =  \rho\, (\tilde{\bf Y}^a)_\rho.
\end{split}
\end{align}
Thus, one can obtain the new metric describing the rotating solution to the vacuum Einstein equations with the same Killing isometries.
As shown previously in Ref.~\cite{Bouchareb:2007ax}, applying this transformation to the five-dimensional static black hole solution (the five-dimensional Schwarzschild-Tangherlini solution~\cite{T-schwarzschild}) generates the five-dimensional rotating black hole solution (the five-dimensional Myers-Perry solution~\cite{Myers:1986un}). 
As will be seen in the following section, however, when applied to vacuum black rings, such as the static black ring and the singly rotating black ring, a Dirac-Misner string singularity inevitably appears on the disk inside the ring, similar to the Harrison transformation studied in Ref.~\cite{Bouchareb:2007ax}.
In the following section, to solve this undesirable problem, we will choose the vacuum rotating black ring having a Dirac-Misner string singularity on the disk inside the ring as the seed of the transformation, as performed in our previous work~\cite{Suzuki:2024coe} for the Harrison transformation.

\section{Seed solution for the Ehlers transformation}\label{sec:ism}

In our previous works~\cite{Suzuki:2023nqf,Suzuki:2024phv}, as a vacuum seed for the Harrison transformation, we chose the vacuum rotating black lens with a horizon cross-section of lens space $L(n;1)$ $(n=0,1,2,\ldots)$~\cite{Chen:2008fa}, which initially has a Dirac-Misner string singularity. It should be noted that in Ref.~\cite{Chen:2008fa}, they removed this singularity, but we did not, since we eliminated it by appropriately adjusting the parameters of the solution after the Harrison transformation.
Now, we follow the same procedure as in Ref.~\cite{Suzuki:2023nqf,Suzuki:2024phv}. 
Furthermore, we use the same seed solution as the seed used when constructing the capped black hole or charged dipole black ring~\cite{Suzuki:2023nqf,Suzuki:2024phv}.

\subsection{Black lens with a Dirac-Misner string singularity as a seed}

The metric for the vacuum seed solution (see Refs.~\cite{Suzuki:2023nqf, Suzuki:2024phv} for details on its construction) that we use for the Ehlers transformation is written in the C-metric form as follows:
\begin{align}
 ds^2 &= -\frac{\bar{H}(y,x)}{\bar{H}(x,y)}(d\bar{t} + \bar{\Omega}_\psi(x,y) d\bar{\psi} + \bar{\Omega}_\phi(x,y) d\bar{\phi})^2
 + \frac{\bar{F}(y,x)}{\bar{H}(y,x)}d\bar{\psi}^2 - \frac{2\bar{J}(x,y)}{\bar{H}(y,x)} d\bar{\psi} d\bar{\phi}- \frac{\bar{F}(x,y)}{\bar{H}(y,x)}d\bar{\phi}^2\nonum
 &+ \frac{\ell^2 \bar{H}(x,y)}{4(1-\gamma)^3(1-\nu)^2(1-a^2)(x-y)^2}\left( \frac{dx^2}{G(x)}-\frac{dy^2}{G(y)}\right),\label{eq:neutral-sol-cmetric}
\end{align}
where
\begin{align}
G(u)& = (1-u^2)(1+\nu u),\label{eq:G}\\
\bar{H}(x,y)& = 
 \left[ \nu  b^2 c_1^2  (1+\nu )^2(\gamma -\nu ) (1-\gamma )  (1-x^2)
+ \nu(1-\gamma )   (\gamma -\nu ) \left(b (1-\nu )^2 (1-x)-2 c_1 (1+\nu  x)\right)^2\right.\nonum
&\hspace{-0.5cm}+ b^2 c_1^2 (1+\nu )^3 (x+1) (\gamma -\nu )^2 \bigr](1+y)^2
-\bigl[d_5  (1-x)^2+d_6(1-x^2)+ 2(1-\nu)^{-1}d_6 (1+x) (1+\nu  x) \bigr](1+y)\nonum
& \hspace{-0.5cm}+ 4 \left(1-a^2\right) (1-\gamma )^3 (1-\nu )^4 (1-x)
 +2c_2^2  (1-\gamma )  (1-\nu )^2   \left(1-x^2\right)-4d_1  (1-\gamma ) (1-\nu)^2 (1+\nu ) (1+x),\label{eq:H1}\\
   %
\bar{F}(x,y) &=\frac{2 \ell^2}{(1-a^2)(x-y)^2} \biggr[
4 \left[\left(1-a^2\right)^2 (y-1) (1-\gamma )^3 (1-\nu )^3-d_1^2(1+y) \right](1+y \nu ) G(x)\nonum
   & +4\left[(1-\nu ) c _2-(1-a b) (\gamma -\nu ) (1+\nu ) c _1\right]^2 (1+x \nu )  (1+x)G(y)\nonum
  &+ \nu^{-1} (1-\nu )^3 (\gamma -\nu ) (d_3^2(1-x^2)G(y) - c _3^2(1-y^2)G(x) )\nonum
& +\frac{G(x)G(y)\left[\left(1-a^2\right) (1-\gamma )
   d_4-(a-b)^2 y (1-\gamma )^2 (\gamma -\nu ) \nu  c _2^2+x (\gamma -\nu ) \nu  \left(b d_1-c _1 c _3\right){}^2\right]}{\nu(1-\gamma) }
\biggr],\\
\bar{J}(x,y)& =  \frac{2\ell^2(1+x)(1+y)}{(1-a^2)(x-y)}
\biggr[4  d_1 
   \left((a-b) (1-\gamma ) (\gamma -\nu ) (1+\nu )-a d_2\right)(1+ \nu x) (1+ \nu y)\nonum
 &  -d_3 c _3 (1-\nu
   )^3 (\gamma -\nu ) (1-x) (1-y)-(a-b) (\gamma -\nu )  c _2 \left(c _1 c _3-b d _1\right)(1-x) (1-y) (1+x \nu ) (1+y \nu )\biggr],
   \end{align}
     \begin{align}
\bar{ \Omega}_\psi(x,y) &= \frac{v_0 \ell (1+y)(1-\nu)} {\nu \bar{H}(y,x)}\biggr[
c _2 \left(c _1 c_3-b d_1\right)(1-x) (1+x \nu ) (1+y \nu ) \nonum
&- (1-\nu )^2 d_2 c _3 (1-x)+(1+x \nu ) d_1 \left(2 \nu(1-a b)  (1-\gamma ) (1+\nu )(1+x)+(1-3   \nu -x (1+\nu )) c _3\right)
\biggr],\\
\bar{ \Omega}_\phi(x,y)& =\frac{v_0 \ell (1+x)}{\bar{H}(y,x)}\biggr[b (1+x) d_1 \left(d_2(1+y) (1+y \nu ) +\nu  c _3\left(1-y^2\right) (1-\nu ) \right)\nonum
&+\frac{2 (a-b) (1-\gamma)^2 \left(2 d_1(1+x \nu )   (1+y \nu )^2 
-(1-\nu )^2 \nu c _3(1-y)  (x+y+\nu +x y \nu ) \right)}{1+\nu }\biggr],\label{eq:Omega_psi}
\end{align}
and the coefficients are given by
\begin{align}\label{eq:def-cfs}
\begin{split}
&v_0 := \sqrt{\frac{2(\gamma^2-\nu^2)}{(1-a^2)(1-\gamma)}},\\
&c _1 :=(1-\gamma) a+(\gamma-\nu) b ,\\
&c _2 := 2 a (1-\gamma ) \nu +b (\gamma -\nu ) (1+\nu ),\\
&c _3 := 2   (1-\gamma ) \nu +b^2 (\gamma -\nu ) (1+\nu ),\\
&d_1 := (\nu +1) c _1^2-(1-\gamma ) (1-\nu )^2,\\
&d_2:= b (\nu +1) c _1 (\gamma -\nu )+2\nu (1-\gamma )   (1-\nu ) ,\\
&  d_3 := \left(a^2-1\right) b (\gamma -1) (\nu +1)-a c _3,\\
&d_4 := b^2 (\gamma -\nu )   \left[(\nu +1)^2 c _1^2 \left(-3 (1-\gamma ) \nu -\nu ^2+1\right)-(1-\gamma ) (1-\nu )^4 (2 \nu +1)\right]\\
&\quad +(1-\gamma   ) \left[\left((1-\nu ) c _2-2 \nu ^2 c _1\right)^2-4 \nu ^2 c _1^2 \left(-\gamma  (\nu +2)+3 \nu
   ^2+1\right)\right],\\
& d_5 := (1-\gamma ) (1-\nu )^3 [(\gamma -3 \nu ) \left(b^2 (\nu -1) (\gamma -\nu )-c_1^2\right)
-2 b c_1 (3 \nu -1) (\gamma -\nu )],\\
& d_6 := c_1(1-\gamma ) (1-\nu) \left(1-\nu ^2\right) \left(c_2-(1-\gamma ) (a-b) (\gamma -\nu )\right).    
   \end{split}
\end{align}
We assume the ranges of the coordinates as
\begin{align}
 -\infty < \bar t < \infty,\quad  -1 \leq x \leq 1,\quad -1/\nu \leq y \leq -1,\label{eq:rangexy}
\end{align}
where, unlike the Harrison transformation studied in our previous works~\cite{Suzuki:2023nqf,Suzuki:2024phv,Suzuki:2024coe}, we do not impose the periodicities at this stage:
\begin{align}
 0 \leq \bar \psi \leq 2\pi,\quad 0\leq \bar\phi \leq 2\pi,\label{eq:rangetpsiphi}
\end{align}
which correspond to the conditions for the absence of conical singularities on $y=-1$ and $x=-1$, respectively. After performing the Ehlers transformation on this seed solution, we will impose the periodicities on the angular coordinates $\psi$ and $\phi$. This solution has five parameters $(\ell,\nu,\gamma,a,b)$, which have the ranges
\begin{align}
\ell>0,\quad  0<\nu <  \gamma < 1, \quad -1<a<1.\label{eq:nugamrange}
\end{align}

The boundaries of the C-metric coordinate $(x,y)$ for the seed solution can be described as follows: 
\begin{enumerate}[(i)]
\item
$\bar \phi$-rotational axis : 
$\partial \Sigma_{\bar \phi}=\{(x,y)|x=-1,-1<y<-1/\nu \}$ with the rod vector 
$v_{\bar \phi}:=(0,0,1)=\partial/\partial\bar\phi$, where it should be noted that in the assumption of the periodicity $\bar\phi\sim \bar\phi +2\pi$ from the coordinate ranges (\ref{eq:rangetpsiphi}) ensures the absence of the conical singularities, but we do not impose this.
\item Horizon: 
$\partial \Sigma_{\cal H}=\{(x,y)|-1<x<1,y=-1/\nu \}$ with the rod vector $
 v_{\cal H} := (1, \bar \omega_\psi, \bar \omega_\phi),
$ 
where
\begin{align}
\begin{split}
&\bar\omega_\psi =\frac{v_0 \left(1-a^2\right) (1-\gamma )}{2 \ell (\gamma +\nu ) \left(1-a^2+a (a-b) \gamma -(1-a b) \nu \right)},\\
&\bar\omega_\phi = \bar\omega_\psi \frac{2 a (1-\gamma ) \nu -\left(1-a^2\right) b (1-\gamma ) (1+\nu )+a b^2 (\gamma -\nu ) (1+\nu )}{2 (1-\gamma ) \nu +b^2 (\gamma -\nu )   (1+\nu )}.\label{eq:vH-n}
   \end{split}
\end{align}

\item
Inner axis: 
$\partial \Sigma_{\rm in}=\{(x,y)|x=1,-1<y<-1/\nu \}$ 
with the rod vector 
\begin{align}
 v_{\rm in}:= (v_{\rm in}^{\bar t},  v_{\rm in}^{\bar\psi} , 1),\label{eq:rodvector-3}
\end{align}
with
\begin{align} 
v_{\rm in}^{\bar t} :=v_0 \ell (a-b),\quad v_{\rm in}^{\bar\psi} := \frac{a d_1+(1-\gamma)(1+\nu)(1-a^2)c_1}{d_1},\label{eq:lensN}
\end{align}
where we note that the presence of the nonzero $t$-component $v_{\rm in}^{\bar t}$ denotes the existence of a the Dirac-Misner string singularity.

\item
$\bar \psi$-rotational axis: 
$\partial \Sigma_{\bar \psi}=\{(x,y)|-1<x<1,y=-1 \}$
with the rod vector $v_{\bar \psi} :=(0,1,0)=\partial/\partial\bar \psi$, 
where the periodicity $\bar\psi\sim \bar\psi +2\pi$ from the coordinate ranges (\ref{eq:rangetpsiphi}) also ensures the absence of the conical singularities, but as mentioned above, we do not impose this.

\item Infinity:  
$\partial \Sigma_\infty =\{(x,y)|x\to y \to -1 \}$.
In the limit $x\to y \to -1\ (r\to\infty)$, the metric (locally) asymptotes to the five-dimensional Minkowski metric 
\begin{align}
ds^2 \simeq -d\bar t^2+dr^2+r^2 d\theta^2 + r^2 \sin^2\theta d\bar\psi^2+r^2\cos^2\theta d\bar\phi^2,
\end{align}
 where the coordinates $(r,\theta)$ are defined by
\begin{align}
 x = -1 + \frac{4 \ell (1-\nu) \cos^2\theta }{r^2},\quad y = -1 - \frac{4\ell(1-\nu) \sin^2\theta}{r^2}.\label{eq:asym-xy}
\end{align}

\end{enumerate}

As previously discussed in Ref.~\cite{Suzuki:2024phv}, by setting $a=b$ where the metric precisely describes the rotating black lens in Ref.~\cite{Chen:2008fa}, the Dirac-Misner string singularity~\cite{Misner:1963fr} on the inner axis $\partial\Sigma_{\rm in}$  can be removed. However, similar to the approach taken for the charged dipole black ring in Ref.~\cite{Suzuki:2024coe} and the capped black hole~\cite{Suzuki:2024phv}, we choose to use the vacuum black lens with a Dirac-Misner string singularity as the seed for the Ehlers transformation. Therefore, we do not assume its absence  before the transformation. 
In the following section, we will eliminate this singularity after the transformation by appropriately adjusting the parameters of the solution.

\subsection{The seed solution in the Euler angles}
The Ehlers transformation~(\ref{eq:sl3r-trans}) requires the choice of either $(\xi_0,\xi_1)=(\partial/\partial t,\partial/\partial \phi_+)$ or $(\xi_0,\xi_1)=(\partial/\partial t,\partial/\partial \phi_-)$, where 
 $\phi_\pm$ are the Euler angles introduced in Sec.~\ref{sec:GStr}. 
In the choice of the Killing vectors $(\xi_0,\xi_1)=(\partial/\partial t,\partial/\partial \phi_+)$ and $\xi_2=\partial/\partial \phi_-$ ($(x^0,x^1,x^2)=(t,\phi_+,\phi_-)$), 
the scalar fields $\Phi^A=\{\lambda_{ab},\omega_a\}$ and the functions $\tau$, $a^a{}_2$ can be expressed as
\begin{align}
\begin{split}
&\lambda_{00} = -\frac{\bar{H}(y,x)}{\bar{H}(x,y)},\quad \lambda_{01} = - \frac{\bar{H}(y,x)}{2 \bar{H}(x,y)}(\bar{\Omega}_\psi(x,y)\pm \bar{\Omega}_\phi(x,y)),\\
& \lambda_{11}=-\frac{\bar{F}(x,y)-\bar{F}(y,x)\pm2\bar{J}(x,y)}{4\bar{H}(y,x)}-\frac{\bar{H}(y,x) (\bar{\Omega}_\phi(x,y)\pm \bar{\Omega}_\psi(x,y))^2}{4 \bar{H}(x,y)},\\
& a_2^0 =  \frac{\bar{F}(x,y) \bar{\Omega}_\psi(x,y)\pm \bar{F}(y,x)\bar{\Omega}_\phi(x,y) \pm \bar{J}(x,y)(\bar{\Omega}_\psi(x,y)\mp \bar{\Omega}_\phi(x,y))}{\bar{F}(x,y)-\bar{F}(y,x)\pm 2\bar{J}(x,y)},\\
& a_2^1 = - \frac{\bar{F}(x,y)+\bar{F}(y,x)}{\bar{F}(x,y)-\bar{F}(y,x)\pm 2\bar{J}(x,y)},\\
& \tau = - \frac{\bar{F}(x,y)-\bar{F}(y,x)\pm 2\bar{J}(x,y)}{4 \bar{H}(x,y)},
\end{split}
\end{align}
and one can find the twist potential is given by\footnote{The Jacobian for the transformation $(\psi,\phi) \to (\phi_+,\phi_-)$ becomes $1/2$. Hence the duality conditions for Eqs.~(\ref{eq:seed-omega}) and (\ref{eq:sol-ai-new-2}) must be given in the rescaled canonical coordinate $(\tilde{\rho},\tilde{z})=(\rho/2,z/2)$.}
\begin{align}
\begin{split}\label{eq:seed-omega}
&\omega_0 = \fr{2}(\bar{\Omega}_\psi(y,x)\pm \bar{\Omega}_\phi(y,x)),\\
&\omega_1 =  -\tau + \frac{2\bar{K}_\pm (x,y)}{\bar{H}(x,y)},
\end{split}
\end{align}
where $\bar{K}_\pm(x,y)$ is a polynomial of $x$ and $y$ given by
\begin{align}
&\bar{K}_\pm (x,y) = \frac{\ell^2}{1\pm a} \biggr[ -2 k^\pm_1 (x+1) \left(y^2-1\right)+k^\pm_2 (x+y-2) (\nu    (-x y+x+y+1)+2)+k^\pm_3 (x+1) (y+1)\nonum
& + k^\pm_4 \left(\nu ^2 \left(x^2 (y-1)+x \left(y^2+7 y+2\right)-y^2+2 y+1\right)+\nu 
   \left(-\left(x^2 (y-1)\right)-x \left(y^2-5\right)+y^2+5 y+6\right)+(x+1) (y+1)\right) \nonum
&   +k^\pm_5   \left(\nu ^2 \left(x^2 (y-1)+x \left(y^2+3\right)-y^2+3 y+2\right)+\nu  \left(x^2 (-y)+x^2-x
   y^2+8 x y+x+y^2+y+6\right)+4 (x+y)\right)\nonum   
&+k^\pm_6 (x+1) (y+1) (x-y)+k^\pm_7
   (x-1) (y-1) (x-y)+k^\pm_8 (\nu  x+1) (x-y) (\nu  y+1)+k^\pm_9 \bar{H}(x,y)\biggr], 
\end{align}
where $\{k^\pm_{i}\}_{i=1,\dots,9}$ are given in the Appendix and $k^+_9$ is the integration constant determined so that the coset matrix has the desirable behavior at the infinity~(\ref{eq:coset-inf}).

\section{Transformed solution}\label{sec:sol}

In what follows, for a simpler expression, we rescale the transformation parameter as $\beta := \ell v_0  \tilde{\beta} /4$.
The new scalar fields  $\Phi^{\prime A}=\{\lambda'_{ab},\omega'_a\}$ and the new  functions $\tau'$, $a^{\prime a}{}_2$  after the transformation~(\ref{eq:sl3r-trans}) can be obtained through Eqs.~(\ref{eq:scalar-trans}) and ~(\ref{eq:sol-ai-new}).
The functions 
 $a^{\prime a}{}_2\ (a=0,1)$ can be expressed into the following lengthy form 
\begin{align}\label{eq:sol-ai-new-2}
a'^a{}_2 &={\cal A}^a_0 + \fr{(x-y)^2(\bar{F}(x,y)-\bar{F}(y,x)+2\bar{J}(x,y))}\biggr[{\cal A}^a_1
   \left(\nu  y \left(x^2-y^2\right)+\nu ^2 x \left(x^2-1\right) y-y^2+1\right)\nonum
 &+{\cal A}^a_2 \left(-\nu  x^3   y^3-x^2 y^3-x y^2+\nu  x y+x+y^3\right)+{\cal A}^a_3 (x-y) \left(\nu  x^2 y^2+\nu  x   y+x+y\right) -{\cal A}^a_4 x^3 \left(y^2-1\right) (\nu  y+1)\nonum
 &-\frac{{\cal A}^a_5 \left(x^2-1\right) y^3 (\nu  x+1)}{\nu }+{\cal A}^a_6
   \left(x^2-1\right) y^2 (\nu  x+1)
     -{\cal A}^a_7 \left(x^2-1\right) y (\nu  x+1)  +\frac{{\cal A}^a_8 \left(x^2-1\right) y^4   (\nu  x+1)}{\nu }
\biggr],
\end{align}
where due to the lengthy expressions, we do not show the coefficients ${\cal A}^a_{k} (k=0,\dots,8)$ explicitly. 
Note that the integration constants ${\cal A}_0^a$ are set so that $a'^a{}_2$ asymptotes to $0+C \cos(2\theta)$ with a constant $C$ as $r\to\infty$ in Eq.~(\ref{eq:asym-xy}). The transformed metric is not at the rest frame but has a global rotation at infinity, which can be removed by redefining the coordinates $(\bar{t},\phi_+,\phi_-) \to (t,\psi,\phi)$.
\begin{align}\label{eq:redefine-coord}
\ \bar{t} = t + {\cal C}(\psi+\phi),\quad \phi_+ = \sqrt{D_0} (\psi+\phi),\quad
 \phi_- = \frac{1}{\sqrt{D_0}}(\psi-\phi),
\end{align}
where $D_0:=D| _{(x,y)=(-1,-1)}$ is given by
\begin{align}
 D_0 := 1 + D_{0,2}\beta^2+D_{0,3}\beta^3+D_{0,4}\beta^4, \label{eq:defD0}
\end{align}
and 
\begin{align}
{\cal C} := \frac{\ell v_0}{8\sqrt{D_0}}\left(2  D_{0,2} \beta +3 D_{0,3} \beta^2+4 D_{0,4} \beta^3\right),
\end{align}
with the coefficients
\begin{align}
\begin{split}
&D_{0,2} = -\frac{6 \left(\left(1-a^2\right) (1-\nu )+(b-a)^2 (\gamma
   -\nu )\right)}{\gamma -\nu },\\
&D_{0,3} = \frac{1}{\gamma^2 -\nu^2 }\biggr[-8 (1-a)^2 (1+a) (1+b)
-8 (a-b)^3 \gamma ^2+8 (1-a)(1-\gamma)
   \left(-3 a+\left(3-a+a^2\right) b+(1-a) b^2\right) \nu\nonum
&\quad   +8   \left(1-3 a b+\left(1+a+a^2\right) b^2-b^3\right) \nu   ^2 -8 (1-a) \gamma \left(a (1+2 a)-\left(1+3 a-a^2\right) b+(1-a) b^2\right)   \biggr],
   \\   
& D_{0,4} = \frac{1}{(\gamma^2 -\nu^2 )^2}\biggr[-3 (a-b)^4 \gamma ^4+4 (1-a)^3 (1+a) (3+2 b) \nu  - (1-a)^2 \left\{3-26 a+7 a^2+8 \left(2-3 a+a^2\right) b+16   b^2\right\} \nu ^2\\
&\quad -2 (1-a) \left\{3-9 a+2 \left(6-5 a+5   a^2\right) b-(1+5 a) b^2-4 (1-a) b^3\right\} \nu ^3\\
&\quad +(1-b)   \left\{-3-3 (1-4 a) b-\left(5-4 a+8 a^2\right) b^2+3   b^3\right\} \nu ^4\\
&\quad+\gamma ^3 \left\{-2 (1-a) (a-b)^2 (1+5
   a+4 (-1+a) b)+2 (-1+a) (-a+b)^2 (3 (-3+a)+4 (-1+a) b) \nu
   \right\}\\
&\quad   +\gamma ^2 \left\{(1-a)^2 \left(1-6 a-11 a^2+8   \left(1+a-2 a^2\right) b+16 a b^2\right)\right.\\
&\quad\quad \left.+2 (1-a) \left(3
   \left(1+a-8 a^2+4 a^3\right)+2 \left(-4+21 a-17 a^2+6
   a^3\right) b+\left(-17+27 a-16 a^2\right) b^2-4 (1-a)
   b^3\right) \nu \right.\\
&\quad\quad   + \nu^2 \left(-7+40 a-60 a^2+40 a^3-7 a^4-4
   \left(6-17 a+24 a^2-9 a^3+2 a^4\right) b\right.\\
&\quad\quad\quad \left.  \left. +2 \left(-7+28 a-11
   a^2+8 a^3\right) b^2-8 \left(1+a+a^2\right) b^3+6
   b^4\right) \right\}\\
&\quad   
   +\gamma  \left\{-4 (1-a)^3 (1+a)
   (1+2 b)-2 (1-a)^3 \left(9+15 a+4 (-1+3 a) b-8 b^2\right)
   \nu \right. \\
&\quad  \quad \left. + 2 (1-a) \left(8-28 a+33 a^2-7 a^3+\left(20-54 a+30
   a^2-8 a^3\right) b+\left(17-19 a+8 a^2\right) b^2
   -4 (1-a)   b^3\right) \nu ^2 \right.\\
&\quad \quad\left. 
   +2 (1-a) \left(3-9 a+2 \left(6-5 a+5
   a^2\right) b-(1+5 a) b^2-4 (1-a) b^3\right) \nu
   ^3\right\}\biggr]. 
\end{split}
\end{align}

In the new coordinates~(\ref{eq:redefine-coord}), one can write the transformed metric again in the C-metric form
\begin{align}
 ds^2 &= -\frac{H(y,x)}{H(x,y)}(dt + \Omega_\psi(x,y) d\psi + \Omega_\phi(x,y) d\phi)^2
 + \frac{F(y,x)}{H(y,x)}d\psi^2 - \frac{2J(x,y)}{H(y,x)} d\psi d\phi- \frac{F(x,y)}{H(y,x)}d\phi^2\nonum
 &+ \frac{\ell^2 H(x,y)}{4(1-\gamma)^3D_0(1-\nu)^2(1-a^2)(x-y)^2}\left( \frac{dx^2}{G(x)}-\frac{dy^2}{G(y)}\right),\label{eq:neutral-sol-cmetric-2}
\end{align}
where each functions are given by
\begin{align}
H(x,y) &=(1+y)^2 \biggr[ \nu  d_1 f_7^2 (1-\nu )  \left(1-x^2\right) \left(2 (\gamma -1) \nu +c_3\right)+ 2 c_3 c_1^2 f_6^2 (\gamma -\nu) (\nu  x+1)^2 \nonum
&\quad  -(\gamma -\nu ) (1-\nu )  \left(c_1^2 c_3 f_6^2-2 (1-\gamma ) \nu  \left(c_1-b (1-\nu   )\right){}^2 f_8^2\right) (1-x) (1+x \nu )\biggr]\nonum
&  -(1+y)  \biggr[ d_5 g_6 (1-x)^2  +d_6 g_7 \left(1-x^2\right)    +2(1-\nu)^{-1} d_6  g_8(1+x)  (1+x \nu )  \biggr] \nonum
& + 4 \left(1-a^2\right) (1-x) (1-\gamma )^3 (1-\nu )^4 D_0-4 (1+x) (1-\gamma ) (1-\nu )^2 (1+\nu ) d_1 f_2^2+2 \left(1-x^2\right) (1-\gamma ) (1-\nu
   )^2 c_2^2 f_5^2,  \\
F(x,y)&=\frac{2\ell^2}{D_0(1-a^2) (x-y)^2}\biggr[
4 (1+y \nu ) G(x) \left\{ \left(1-a^2\right)^2 (-1+y) (1-\gamma )^3 (1-\nu )^3 D_0^2-(1+y) d_1^2 f_1^2 f_2^2\right\}\nonum
&+4 (1+x) (1+x \nu )  G(y) \left((1-a b) (\gamma -\nu ) (1+\nu ) c_1 -(1-\nu ) c_2 \right){}^2 g_{5}^2\nonum
&+ \nu^{-1} (1-\nu )^3 (\gamma -\nu ) \biggr\{\left(y^2-1\right) G(x) c_3^2 f_3^2 f_4^2+\left(1-x^2\right) G(y) d_3^2 g_2^2\biggr\}\nonum
& + G(x) G(y) \biggr\{\frac{x (\gamma -\nu ) (c_1 c_3f_3 f_6-b d_1f_1f_7)^2 }{1-\gamma }-(a-b)^2 y (1-\gamma ) (\gamma -\nu ) c_2^2
   f_5^2 f_8^2+\frac{(1-a^2) d_4 D_0 g_4}{\nu }\biggr\}   
\biggr],\\
J(x,y)& =  \frac{2 \ell^2(1+x)(1+y)}{D_0(1-a^2)(x-y)}
\biggr[4 d_1 f_1 f_2 g_{5} \biggr\{(1-a b) (\gamma -\nu ) (1+\nu ) c_1 -(1-\nu ) c_2 \biggr\} (1+x \nu ) (1+y \nu ) \nonum
& - (1-x) (1-y) (\gamma -\nu ) \biggr\{(a-b) (1+x \nu ) (1+y \nu ) c_2 f_5 f_8 \left(c_1 c_3 f_3 f_6 -b d_1 f_1 f_7 \right) 
+(1-\nu )^3 c_3   d_3 f_3 f_4 g_2\biggr\}
\biggr],\\
\Omega_\psi (x,y)&= \frac{\ell v_0 (1+y)}{\sqrt{D_0} H(y,x)}\biggr[
\nu^{-1}c_2 f_5 \left(c_1 c_3 f_4 f_{16}-b d_1 f_2 f_{13}\right)  (1-x) (1-\nu ) (1+x \nu ) (1+y \nu ) \nonum
&+ (1+y \nu ) \left\{(1+x) (1+x \nu ) c_1 c_3 f_2 f_9 f_{10}-\frac{1}{2} b \left(1-x^2\right) (\gamma -\nu ) (1-\nu ) c_3 f_4 f_{10} f_{12}\right\}\nonum
&-2 (1+x) (1-\nu ) (1+x \nu ) d_1 f_2 f_{11}  f_{15} \left(c_3-(1-a b) (1-\gamma ) (1+\nu ) \right)\nonum
&+\nu^{-1} c_3 \left(d_1-d_2\right) f_4 g_3(1-x) (1-\nu )^2 (1+x \nu ) +c_3 d_2 f_4 g_{10} (1-\nu )^2\left(1-x^2\right) 
 \biggr],\\
\Omega_\phi(x,y) &= \frac{\ell v_0 (1+x)}{\sqrt{D_0}H(y,x)}\biggr[
b (1+x) (1+y) (1+y \nu ) d_1 d_2 f_1 f_{13} f_{14}\nonum
& +\frac{(1+x) \left( 1-y^2\right) (1-\nu ) \nu  c_3 \left(b (1+\nu ) d_1-2 (a-b) (1-\gamma )^2 (1-\nu ) \nu \right) f_3 g_{11}}{1+\nu }\nonum
&+\frac{(y-1) (1-\gamma ) (1-\nu ) \nu  (1+y \nu + x (1+y (4-3 \nu ))) c_3 f_1 f_3 f_{10}}{1+\nu }\nonum
&+\frac{2 (a-b) (1-\gamma )^2 \left((y-1) (x+y) (1-\nu )^3 \nu  c_3 f_3 g_{9}+2 (1+x \nu ) (1+y \nu )^2 d_1 f_1 g_{1}\right)}{1+\nu }
\biggr],
\end{align}
where $\{f_i\}_{i=1,\dots,16}$ and $\{g_i\}_{i=1,\dots,11}$ represent the polynomials in $\beta$ up to $\ord{\beta^2}$ and $\ord{\beta^4}$, respectively, with coefficients that are functions of the parameters $(a,b,\gamma,\nu)$. 
The explicit expressions for $f_i$ and $g_i$ can be seen  in the Appendix~\ref{sec:fi-gi}.
Both $f_i$ and $g_i$ are defined such that $f_i(\beta=0)=1$ and $g_i(\beta=0)=1$, except $f_{10}$ which vanishes at $\beta=0$.
Note that $D_0$ defined in Eq.~(\ref{eq:defD0}) has a simpler expression in terms of $f_i$
\begin{align}\label{eq:defD0-simp}
 D_0 = f_8^2-\frac{d_1 \left(f_1^2-f_8^2\right)}{\left(1-a^2\right) (1-\gamma ) (1-\nu^2 ) }+\frac{c_3
   \left(f_3^2-f_8^2\right)}{\left(1-a^2\right) (1-\gamma ) (1+\nu )}.
\end{align}

\subsection{Asymptotic infinity and the rod structure}
 The boundaries for the transformed solution in the C-metric coordinates, including the infinity, can be summarized as follows:
\begin{enumerate}[(i)]
\item
$\phi$-rotational axis : 
$\partial \Sigma_\phi=\{(x,y)|x=-1,-1<y<-1/\nu \}$ with the rod vector 
$v_\phi:=(0,0,1)=\partial/\partial \phi$, where  we impose the periodicity of the coordinate $\phi$ as $\phi\sim \phi +2\pi$ to ensures the absence of the conical singularities on $\partial \Sigma_\phi$, 
\item Horizon: 
$\partial \Sigma_{\cal H}=\{(x,y)|-1<x<1,y=-1/\nu \}$ with the rod vector $
 v_{\cal H} := (1, \omega_\psi, \omega_\phi),
$ 
where
\begin{align}
\omega_\psi= \bar{\omega}_\psi\frac{\sqrt{D_0} f_4}{g_3},\quad \omega_\phi = \bar{\omega}_\phi \frac{\sqrt{D_0} g_2 }{f_3 g_3},
\end{align}

\item
Inner axis: 
$\partial \Sigma_{\rm in}=\{(x,y)|x=1,-1<y<-1/\nu \}$ 
with the rod vector 
\begin{align}
 v_{\rm in}=(v_{\rm in}^t,v_{\rm in}^\psi,1)
\end{align}
where
\begin{align}
v_{\rm in}^t &=\frac{\ell v_0 (a-b) \tilde{g}_1}{\sqrt{D_0} f_1} =\frac{\ell v_0 (a-b+\delta \tilde{g_1})}{\sqrt{D_0} f_1},\\
 v_{\rm in}^\psi &=\frac{D_0}{f_1f_2}(v_{\rm in}^{\bar{\psi}}-1)+1,
\end{align}
and $\tilde{g_1}$ and $\delta \tilde{g_1}$ are defined by, respectively, 
\begin{align}
\tilde{g_1} := 1 + \frac{\delta \tilde{g}_1}{a-b} := g_{1}+\frac{\nu c_3 f_3 f_{10} }{(1-\gamma ) d_1 (\nu +1) (a-b)}= f_{13} f_{14}+ \frac{c_3 f_3 f_{10}}{2 (a-b) (1-\gamma ) d_1}-\frac{c_1 c_3\left(f_{13} f_{14}-f_{16}   f_{19}\right)}{2 (a-b) (1-\gamma )^2 \nu }.
\end{align}
Here, we should note that the presence of the $t$-component denotes the existence of a Dirac-Misner string singularity.

\item
$\psi$-rotational axis: 
$\partial \Sigma_\psi=\{(x,y)|-1<x<1,y=-1 \}$
with the rod vector $v_\psi :=(0,1,0)=\partial/\partial \psi$, where 
we impose the periodicity of the coordinate $\psi$ as $\psi\sim \psi +2\pi$ to ensures the absence of the conical singularities on $\partial \Sigma_\psi$,

\item Infinity:  
$\partial \Sigma_\infty =\{(x,y)|x\to y \to -1 \}$:  
Introducing the coordinates $(r,\theta)$ by
\begin{align}
 x = -1 + \frac{4 \ell (1-\nu) \cos^2\theta }{r^2},\quad y = -1 - \frac{4\ell(1-\nu) \sin^2\theta}{r^2},\label{eq:asym-xy}
\end{align}
we find that the limit $x\to y \to -1\ (r\to\infty)$ corresponds to the asymptotic infinity, since 
in this limit, the metric behaves as the five-dimensional Minkowski metric 
\begin{align}
ds^2 = -\left(1-\frac{8G_5 M}{3\pi r^2}\right)dt^2-\frac{8G_5 J_\psi \sin^2\theta}{\pi r^2}dt d\psi-\frac{8G_5 J_\phi \cos^2\theta}{\pi r^2}dt d\phi+dr^2 + r^2\sin^2\theta d\psi^2+r^2 \cos^2\theta d\phi^2+r^2d\theta^2,
\end{align}
where the mass $M$ and two angular momenta $J_\psi$, $J_\phi$ can be written as
\begin{align}
 M &=  \frac{3 \ell^2 \pi  (\gamma +\nu ) \left(d_1-(1-\nu ) c_3\right)}{4 (1-a^2) (1-\gamma )^2 (1+\nu )D_0} \biggr[
f_8^2+\frac{(1+\nu ) c_2^2 \left(f_8^2-f_5^2\right)}{(1-\nu ) (\gamma +\nu ) \left(d_1-(1-\nu )   c_3\right)}\nonum
   &\quad +\frac{(1+\nu ) d_1 \left((1-\gamma ) \left(f_8^2-f_1^2\right)-(1+\nu )   \left(f_8^2-f_2^2\right)\right)}
    {(1-\nu ) (\gamma +\nu ) \left(d_1-(1-\nu ) c_3\right)} \biggr], \label{eq:mass}\\
  J_\psi &= -\frac{\ell^3 \pi v_0 \left(c_3 f_4 g_3\left(d_1-d_2\right) +c_2 f_5 \left(c_1 c_3 f_4 f_{16}-b d_1 f_2 f_{13}\right)\right)}
  {4 D_0^{3/2} \nu\left(1-a^2\right)  (1-\gamma
   )^3  }, \label{eq:Jpsi-trans}\\
  J_\phi &= -\frac{\ell^3 \pi  v_0 \left(  2 \nu  c_3 f_3 f_1 f_{10}-(2 \nu  c_3 f_3 g_{9}+ d_1 f_1   g_{1})(a-b) (1-\gamma ) (1-\nu )\right)}{2D_0^{3/2} \left(1-a^2\right)  (1-\gamma )^2 (1-\nu^2 ) },\label{eq:Jphi-trans}
\end{align}
where we used the expression of Eq.~(\ref{eq:defD0-simp}) in the mass formula.

\end{enumerate}

Finally, we summarize some observations on how the Ehlers transformation changes the rod structure. 
This transformation does not change the positions of both the rotational axes at $x=-1$ and at $y=-1$. Additionally, it does not alter the directions of the rod vectors after the removal of the global rotation by the coordinate transformation. 
However, it does not preserve the regularity, specifically the absence of conical singularities on the two axes. 
Therefore, to ensure the regularity of the transformed solution, we did not impose regularity on the seed solution.
Additionally, this transformation also does not change the positions of the horizon $\partial \Sigma_{\cal H}$ at $y=-1/\nu$ and the inner axis $\partial \Sigma_{\rm in}$ at $x=1$, but it changes the directions of the rod vectors. Specifically, $v_{\cal H}$ and $v_{\rm in}$ change from $(1,\bar\omega_\psi,\bar\omega_\phi)$ to $(1,\omega_\psi,\omega_\phi)$ and from $(v_{\rm in}^{\bar t},v_{\rm in}^{\bar \psi},1)$ to $(v_{\rm in}^{t},v_{\rm in}^{\psi},1) = ((v_{\rm in}^{\bar t}+\ell v_0 \delta \tilde g_1)D_0^{-\frac{1}{2}}f_1^{-1}, D_0 (v_{\rm in}^{\bar{\psi}}-1)/(f_1f_2)+1,1)$, respectively.

Since the presence of the nonzero $t$-component $v_{\rm in}^t$ in $v_{\rm in}$ still yields a Dirac-Misner string singularity, one can remove it by setting
\begin{align}
a-b+\delta \tilde g_1=0, 
\end{align}
where note that $\delta \tilde g_1\to 0$ as $\beta \to 0$. Therefore, if we choose the seed solution with $\beta=0$ that does not have the Dirac-Misner string singularity (corresponding to $a=b$), as in Ref.~\cite{Bouchareb:2007ax}, the Dirac-Misner string singularity inevitably appears due to $\delta\tilde g_1 \not=0$ for $\beta \not=0$. Thus, to obtain a regular solution without a Dirac-Misner string singularity by the transformation, we must choose a singular seed solution having a Dirac-Misner string singularity. Moreover, since $v_{\rm in}^{\psi}\not= v_{\rm in}^{\bar \psi}$ for $\beta\not= 0$, this transformation does not preserve the horizon topology, which implies that even if the horizon cross-section of the seed solution has a ring topology $S^2\times S^1$ ($v_{\rm in}^{\bar\psi}=0$), the transformed solution does not ($v_{\rm in}^{\psi}\not=0$).

\subsection{Regularity conditions for black rings}
Now, we consider the condition that the transformed solution describes a regular black ring, meaning that the spacetime has neither curvature singularities, conical singularities, orbifold singularities, nor Dirac-Misner string singularities either inside or outside the horizons.

\medskip
As shown in Ref.~\cite{Hollands:2007aj}, the topology of the horizon cross-section is assured to be $S^2\times S^1$ when the rod vector on $\partial\Sigma_{\rm in}$ can be expressed in the form $v_{\rm in}=(0,1,0)$. This requires that 
\begin{align}
v_{\rm in}^t&= \frac{\ell v_0 (a-b+\delta \tilde{g}_1)}{\sqrt{D_0} f_1}= 0,\label{eq:no-dms}\\
v_{\rm in}^\psi &=\frac{D_0}{f_1f_2} (v_{\rm in}^{\bar{\psi}}-1) +1 =0,\label{eq:ring-topology}
\end{align}
where Eq.~(\ref{eq:no-dms}) is the condition for the absence of a Dirac-Mister string singularity on $\partial \Sigma_{\rm in}$ and Eq.~(\ref{eq:ring-topology}) assures the ring topology of the horizon cross-section.
Moreover, the absence of conical singularities on $\partial \Sigma_{\rm in}$ requires the condition:
\begin{align}
\lim_{\rho\to0}\sqrt{\frac{\rho^2 g_{\rho\rho}}{g_{ij}v_{\rm in}^iv_{\rm in}^j}} =\frac{\Delta \phi}{2\pi}\Longleftrightarrow  \frac{d_1^2 f_1^2 f_2^2}{\left(1-a^2\right)^2 (1-\gamma )^3 (1-\nu )^2 (1+\nu ) D_0^2}=1. \label{eq:conifree}
\end{align}
From these three conditions,  the number of the independent parameters are reduced from six $(\ell, a,b,\gamma,\nu,\beta)$ to three. 
Moreover, as discussed in Section IV.A of Ref.~\cite{Suzuki:2024coe}, since the Kretschmann invariant can be written as $R_{\mu\nu\rho\sigma}R^{\mu\nu\rho\sigma}\propto H^{-6}(x,y)$ and $H(x,y)>0$ at infinity $x\to y \to -1$,  
the absence of naked curvature singularities necessitates $H(x,y)>0$ for all $(x,y)$ in the entire region $\{ (x,y)| -1 \leq x \leq 1, -1/\nu \leq y \leq -1\}$.
Therefore, the regular solution must satisfy this at the point $(x,y)=(1,-1)$, 
\begin{align}
&H(x=1,y=-1) = -8(1-\gamma)(1-\nu)^2(1+\nu)d_1 f_2^2 >0  \quad \Longleftrightarrow  \quad d_1<0.\label{eq:necessary-2}
\end{align}

\medskip
The two conditions, (\ref{eq:ring-topology}) and (\ref{eq:conifree}),   lead to
 the $\beta$-independent conditions:
 \begin{align}
\frac{d_1^2(1-v_{\rm in}^{\bar\psi})^2}{(1-\gamma)^3(1-\nu)^2(1+\nu)(1-a^2)^2} 
= \frac{(d_1+c_1(1-\gamma)(1+\nu)(1+a))^2}{(1-\gamma)^3(1-\nu)^2(1+\nu)(1+a)^2}= 1.\label{eq:conifree+ring}
\end{align}
To address this condition efficiently, it is useful to introduce the following parameters, 
\begin{align}
 \tilde{\gamma} := \sqrt{1-\gamma} ,\quad \tilde{\nu} := \sqrt{1+\nu},\quad (0<\tilde{\gamma}<1<\tilde{\nu}<\sqrt{2},\quad \tilde{\gamma}^2+\tilde{\nu}^2<2),
\end{align}
or conversely,
\begin{align}
\gamma = 1-\tilde{\gamma}^2,\quad \nu = \tilde{\nu}^2-1.
\end{align}
In terms of $\tilde{\nu}$ and $\tilde{\gamma}$, Eq.~(\ref{eq:conifree+ring}) presents four distinct branches for $b$:
\begin{align}
 b =\frac{\tilde{\gamma}(\tilde{\gamma}\tilde{\nu}+2-\tilde{\nu}^2)}{\tilde{\nu}(2-\tilde{\gamma}^2-\tilde{\nu}^2)},\quad \frac{\tilde{\gamma}(\tilde{\gamma}\tilde{\nu} - (2-\tilde{\nu}^2))}{\tilde{\nu}(2-\tilde{\gamma}^2-\tilde{\nu}^2)},\quad
 \frac{\tilde{\gamma}(a \tilde{\gamma}\tilde{\nu} + 2-\tilde{\nu}^2)}{\tilde{\nu}(2-\tilde{\gamma}^2-\tilde{\nu}^2)},\quad  \frac{\tilde{\gamma}(a \tilde{\gamma}\tilde{\nu} - (2-\tilde{\nu}^2))}{\tilde{\nu}(2-\tilde{\gamma}^2-\tilde{\nu}^2)}.\label{eq:n0sol-b}
 \end{align}
 The latter two of the four branches are not regular, as they result in $d_1=0$. 
The first two branches are interconnected through the transformation $(\tilde{\nu},\tilde{\gamma}) \to (-\tilde{\nu},\tilde{\gamma})$ or $(\tilde{\nu},\tilde{\gamma})\to (\tilde{\nu},-\tilde{\gamma})$, 
and therefore can be merged to a single branch
\begin{align}\label{eq:n0sol-b1}
b =\frac{\tilde{\gamma}(\tilde{\gamma}\tilde{\nu}+2-\tilde{\nu}^2)}{\tilde{\nu}(2-\tilde{\gamma}^2-\tilde{\nu}^2)},
\end{align}
by letting $\tilde{\nu}$ and $\tilde{\gamma}$ to take the negative values
\begin{align}\label{eq:range-tnu-tgam}
 -\sqrt{2} < \tilde{\nu} <-1 <\tilde{\gamma} <1, \quad  \tilde{\gamma}^2+\tilde{\nu}^2<2 \quad {\rm or} \quad -1<\tilde{\gamma} <1 < \tilde{\nu} < \sqrt{2}, \quad  \tilde{\gamma}^2+\tilde{\nu}^2<2.
\end{align}

Substituting Eq.~(\ref{eq:n0sol-b1}) into Eq.~(\ref{eq:ring-topology}) results in the product of two quadratic equations in $\beta$ as follows: 
\begin{align}\label{eq:n0eq-beta}
&\biggr[\tilde{\nu }^2 \left(2-\tilde{\gamma }^2-\tilde{\nu }^2\right)^2 \left(\tilde{\nu
   }^2+\tilde{\gamma } \tilde{\nu }-2\right)-2 \beta  \tilde{\nu } \left(2-\tilde{\gamma
   }^2-\tilde{\nu }^2\right) \left((1+a) \tilde{\gamma }^3 \tilde{\nu }^2+\tilde{\gamma }
   \left(2-\tilde{\nu }^2\right) \left(2-3 \tilde{\gamma } \tilde{\nu }-(1+a) \tilde{\nu
   }^2\right)+\tilde{\nu } \left(2-\tilde{\nu }^2\right)^2\right)\nonum
   &\quad +\beta ^2 \left((1+a)^2
   \tilde{\gamma }^5 \tilde{\nu }^3+\tilde{\gamma }^3 \tilde{\nu } \left(2-\tilde{\nu
   }^2\right) \left(10+\left(-5-4 a+a^2\right) \tilde{\gamma } \tilde{\nu }-(2+a)^2
   \tilde{\nu }^2\right)\right.\nonum
&\quad\quad \left.    -\tilde{\nu }^2 \left(2-\tilde{\nu }^2\right)^3-\tilde{\gamma }
   \left(2-\tilde{\nu }^2\right)^2 \left(2 \tilde{\gamma }+4 \tilde{\nu }-\left(4+4
   a-a^2\right) \tilde{\gamma } \tilde{\nu }^2-(1+2 a) \tilde{\nu }^3\right)\right)\biggr] \nonum
&\times \biggr[\tilde{\nu }^2 \left(2-\tilde{\gamma }^2-\tilde{\nu }^2\right)^2
-2 \beta  \tilde{\nu }   \left(2-\tilde{\gamma }^2-\tilde{\nu }^2\right) \left((1+a) \tilde{\gamma }^2
   \tilde{\nu }-(\tilde{\gamma }+\tilde{\nu}) \left(2-\tilde{\nu }^2\right)\right)\nonum
&\quad   +\beta ^2 \left((1+a)^2 \tilde{\gamma }^4
   \tilde{\nu }^2-2 (1+a) \tilde{\gamma }^3 \tilde{\nu } \left(2-\tilde{\nu }^2\right)
   +\tilde{\nu } (\tilde{\nu}+2   \tilde{\gamma } )\left(2-\tilde{\nu }^2\right)^2
  +\tilde{\gamma }^2 \left(2-\tilde{\nu }^2\right)
   \left(2-\left(4+a^2\right) \tilde{\nu }^2\right)\right)\biggr]=0.
\end{align}
This has two branches: one is given by
\begin{align}\label{eq:n0sol-beta1}
&\beta=\frac{\tilde{\nu } \left(2-\tilde{\gamma }^2-\tilde{\nu }^2\right)}{{\cal B}_0}\biggr[(1+a) \tilde{\gamma }^3 \tilde{\nu }^2+\tilde{\nu } \left(2-\tilde{\nu   }^2\right)^2+\tilde{\gamma } \left(2-\tilde{\nu }^2\right) \left(2-3 \tilde{\gamma }   \tilde{\nu }-(1+a) \tilde{\nu }^2\right)\nonum
&\quad \quad - (1-a) \tilde{\gamma } \tilde{\nu }
   \sqrt{2\left(2-\tilde{\gamma }^2-\tilde{\nu }^2\right) \left(-2+3 \tilde{\nu
   }^2-\tilde{\nu }^4\right)}\biggr],
\end{align}
with 
\begin{align}
&\quad  {\cal B}_0 = (1+a)^2 \tilde{\gamma }^5 \tilde{\nu }^3-\tilde{\nu }^2 \left(2-\tilde{\nu
   }^2\right)^3+\tilde{\gamma }^3 \tilde{\nu } \left(2-\tilde{\nu }^2\right)   \left(10-\left(5+4 a-a^2\right) \tilde{\gamma } \tilde{\nu }   -(2+a)^2 \tilde{\nu   }^2\right)\nonum
&\quad\quad+\tilde{\gamma } \left(2-\tilde{\nu }^2\right)^2 \left(-2 \tilde{\gamma }-4
   \tilde{\nu }+\left(4+4 a-a^2\right) \tilde{\gamma } \tilde{\nu }^2+(1+2 a) \tilde{\nu
   }^3\right),   
\end{align}
and the other is given by
\begin{align}\label{eq:n0sol-beta2}
\beta =\frac{\tilde{\nu } \left(2-\tilde{\gamma }^2-\tilde{\nu }^2\right)
   \left(\left(\tilde{\gamma }+\tilde{\nu }\right) \left(2-\tilde{\nu   }^2\right)-\tilde{\gamma } \tilde{\nu } \left((1+a) \tilde{\gamma }-(1-a)
   \sqrt{2-\tilde{\nu }^2}\right)\right)}{(1+a)^2 \tilde{\gamma }^4 \tilde{\nu }^2-2
   (1+a) \tilde{\gamma }^3 \tilde{\nu } \left(2-\tilde{\nu }^2\right)+2 \tilde{\gamma }
   \tilde{\nu } \left(2-\tilde{\nu }^2\right)^2+\tilde{\nu }^2 \left(2-\tilde{\nu   }^2\right)^2+\tilde{\gamma }^2 \left(2-\tilde{\nu }^2\right)   \left(2-\left(4+a^2\right) \tilde{\nu }^2\right)}.
\end{align}
Note that Eq.~(\ref{eq:n0eq-beta}) admits two additional branches, which are derived by reversing the sign of the square root terms in Eqs.~(\ref{eq:n0sol-beta1}) and (\ref{eq:n0sol-beta2}).
However, these latter two branches can be combined to the former two through the transformation $(\tilde{\nu} ,\tilde{\gamma}) \to (-\tilde{\nu},-\tilde{\gamma})$.
Therefore,  within the range specified in Eq.~(\ref{eq:range-tnu-tgam}), the solutions for Eqs.~(\ref{eq:ring-topology}) and (\ref{eq:conifree}) are provided by Eqs.~(\ref{eq:n0sol-b1}) and (\ref{eq:n0sol-beta1}) or Eqs.~(\ref{eq:n0sol-b1})  and (\ref{eq:n0sol-beta2}).

\medskip
Next, we consider the condition for the absence of the Dirac-Misner string singularity~(\ref{eq:no-dms}). 
Using Eq.~(\ref{eq:n0sol-b1}) with Eq.~(\ref{eq:n0sol-beta1}), we obtain
\begin{align}\label{eq:n0sol-a1}
a = \frac{\left(2-\tilde{\gamma } \tilde{\nu }-\tilde{\nu }^2\right) \left(2 \tilde{\nu
   }-\tilde{\gamma } \left(2-\tilde{\nu }^2\right)\right)}{\tilde{\gamma } \tilde{\nu }
   \left(\tilde{\nu }^3-\tilde{\gamma } \left(2-\tilde{\nu }^2\right)\right)},
\end{align}
and with Eq.~(\ref{eq:n0sol-beta2}),
\begin{align}\label{eq:n0sol-a2}
a = \frac{\tilde{\gamma }^3 \tilde{\nu }^2-\tilde{\gamma }^2 \tilde{\nu } \left(2-\tilde{\nu
   }^2\right)-\left(\tilde{\gamma }-\tilde{\nu }\right) \left(2-\tilde{\nu
   }^2\right)^2+\sqrt{2-\tilde{\nu }^2} \left(\tilde{\gamma }^2-\tilde{\nu }^2\right)
   \left(2-\tilde{\nu }^2\right)}{\tilde{\gamma } \tilde{\nu }^2 \left(2-\tilde{\gamma
   }^2-\tilde{\nu }^2\right)},
\end{align}
where the former and the latter  lead to, respectively, 
\begin{align}
d_1 = -\frac{4 \tilde{\gamma }^2 \left(2-\tilde{\nu }^2\right)^2 \left(\tilde{\nu }^2-1\right)
   \left(\tilde{\nu }^2-\tilde{\gamma }^2\right)}{\left(\tilde{\nu }^3-\tilde{\gamma }
   \left(2-\tilde{\nu }^2\right)\right)^2}<0,
   \end{align}
and 
\begin{align}
d_1=\frac{2 \tilde{\gamma }^2 \left(2-\tilde{\nu }^2\right)^2 \left(\tilde{\nu }^2-1\right)
   \left(\tilde{\nu }^2-\tilde{\gamma }^2\right) \left(\tilde{\gamma}-\sqrt{2-\tilde{\nu }^2}\right)^2}{\tilde{\nu }^2 \left(2-\tilde{\gamma   }^2-\tilde{\nu }^2\right)^2}>0.
\end{align}
Thus, the most general solution subject to the three conditions~(\ref{eq:no-dms}), (\ref{eq:ring-topology}) and (\ref{eq:conifree}), together with the necessary condition~(\ref{eq:necessary-2}), is given by Eqs.~(\ref{eq:n0sol-b1}), (\ref{eq:n0sol-beta1}) and (\ref{eq:n0sol-a1}) for $(\tilde{\nu},\tilde{\gamma})$ in the range~(\ref{eq:range-tnu-tgam}).
The range~(\ref{eq:range-tnu-tgam}) is covered by the original parameter $(\nu,\gamma)$, if we introduce $\epsilon_1 = \pm 1$ and $\epsilon_2=\pm1$, as
\begin{align}
\tilde{\nu} = \epsilon_1 \sqrt{1+\nu},\quad \tilde{\gamma}= \epsilon_2 \sqrt{1-\gamma},
\end{align}
with which Eqs.~(\ref{eq:n0sol-a1}), (\ref{eq:n0sol-b1}) and (\ref{eq:n0sol-beta1}) are rewritten as
\begin{align}
a &= \frac{(\gamma -3) (1-\nu ) \sqrt{\nu +1}+\epsilon_1 \epsilon_2 \sqrt{1-\gamma } \left(\nu ^2+3\right)}{(1-\gamma ) (1-\nu ) \sqrt{\nu +1}-\epsilon_1\epsilon_2 \sqrt{1-\gamma } (\nu +1)^2},\label{eq:a-exp-2}\\
b &= \frac{\sqrt{1+\nu}(1-\gamma )- \epsilon_1 \epsilon_2 \sqrt{1-\gamma } (1-\nu)}{(\gamma -\nu)\sqrt{1+\nu}} .\label{eq:br-exp-b}\\ 
\beta &= \frac{1}{{\cal B}_1 + \epsilon_1 \epsilon_2 {\cal B}_2\sqrt{(1-\gamma ) (1+\nu )} }\biggr[
\left(1-\nu ^2\right) \left(\gamma ^3 \left(-1+4 \nu -3 \nu ^2\right)+\gamma ^2
   \left(8-3 \nu +4 \nu ^2+7 \nu ^3\right)\right. \nonum
&\quad \left.   +\gamma  \left(-8-8 \nu +\nu ^2-12 \nu ^3-5
   \nu ^4\right)+\nu  \left(8+3 \nu ^2+4 \nu ^3+\nu ^4\right)\right)\nonum
&  + \epsilon_1 \epsilon_2 2 \sqrt{1+\nu } \sqrt{1-\gamma } (1-\nu ) \left(\gamma ^2 \left(-2+\nu +2 \nu ^2+3
   \nu ^3\right)-4 \gamma  \left(-1-\nu +\nu ^3+\nu ^4\right)+\nu  \left(-4-2 \nu -\nu
   ^2+2 \nu ^3+\nu ^4\right)\right)\nonum
&   +2 (\gamma -\nu ) \sqrt{2 (1-\nu ) \nu  (\gamma -\nu )} \left(\epsilon_1\sqrt{1+\nu } (1-\nu )
   \left(4+5 \nu +3 \nu ^2-\gamma  \left(3+3 \nu +2 \nu ^2\right)\right)\right. \nonum
&\quad   \left.  
\epsilon_2   \sqrt{1-\gamma } (1+\nu ) \left(-4+\gamma 
 (1-\nu )^2+\nu -4 \nu ^2-\nu
   ^3\right)\right)
\biggr],\label{eq:br-exp-beta}
\end{align}
where
\begin{align}
&{\cal B}_1 = (1-\nu )^2 \left(-16-12 \nu -37 \nu ^2-29 \nu ^3-3 \nu ^4+\nu ^5
-\gamma ^2 \left(5+17   \nu +11 \nu ^2-\nu ^3\right)+2 \gamma  \left(10+19 \nu +25 \nu ^2+9 \nu ^3+\nu
   ^4\right)\right),\nonum
&{\cal B}_2 =(1-\nu ) \left(16-12 \nu +33 \nu ^2+5 \nu ^3+23 \nu ^4-\nu
   ^5-\gamma ^2 (-1+\nu ) (1+\nu )^2-2 \gamma  \left(6+7 \nu +13 \nu ^2+5 \nu ^3+\nu
   ^4\right)\right).
\end{align}

\section{Equivalence with the Pomeransky-Sen'kov solution}\label{sec:PS0}
To summarize, it can be seen from Eqs.~(\ref{eq:a-exp-2}), (\ref{eq:br-exp-b}) and (\ref{eq:br-exp-beta}) that the transformed solution can be expressed only in terms of the three parameters $(\ell,\nu,\gamma)$. 
As explaned in the previous section, we have the four branches based on the choice of signatures $(\epsilon_1,\epsilon_2)=(\pm 1,\pm 1)$, which we respectively refer to as the $(\pm,\pm)$-branch.
We now demonstrate that each of four branches is identical to the Pomeransky-Sen'kov solution~\cite{Pomeransky:2006bd}, described by the metric~(\ref{eq:metricPS}) and characterized by a three-parameter family $(k,\alpha_1,\alpha_2)$.
This solution describes a doubly rotating black ring with $J_\psi>0$ and $J_\phi>0$ for the parameter range of $k>0$ and $0\leq \alpha_2 \leq \alpha_1<1$.
The parameter region is bounded by the singly rotating solution, i.e., Emparan-Reall black ring solution, $\{(\alpha_1,\alpha_2)| 0<\alpha_1<1,\alpha_2=0\}$ and the extremal solution $\{(\alpha_1,\alpha_2)| 0<\alpha_1=\alpha_2<1\}$.
One can show the equivalence of each branch to the Pomeransky-Sen'kov solution almost in parallel, it suffices to prove it for a single branch, say $(++)$-branch.

\medskip
According to the uniqueness theorem for a vacuum black ring solution~\cite{Hollands:2007aj,Morisawa:2007di,Tomizawa:2009tb}, under the assumptions of the presence of three mutually commuting Killing vector fields, a timelike Killing vector field and two axial Killing vector fields, besides the trivial topology of (${\Bbb R} \times \{ {\Bbb R}^4 \setminus  D^2\times S^2 \} $) 
of the domain of outer communication, an asymptotically flat, stationary rotating black ring with non-degenerate connected event horizon of the cross-section topology $S^2\times S^1$ is characterized uniquely by the mass, two independent angular momenta, and additional information on the rod structure such as the ratio of the $S^2$ radius to the $S^1$.
Therefore, if our transformed solution exactly coincides with the Pomeransky-Sen'kov solution, using this theorem, we can extract the relation between two sets of the parameters, $(\ell,\nu,\gamma)$ in our transformed solution and $(k,\alpha_1,\alpha_2)$ in the Pomeransky-Sen'kov solution by comparing the rod length and asymptotic charges between two solutions. 
To relate these two three-parameter families of solutions, it is sufficient to consider only the rod data (the rod lengths of the horizon and inner axis) and the mass for the two solutions. 
In the following discussion, we examine the correspondence between our solution obtained in the previous section and the non-extremal Pomeransky-Sen'kov solution, whose parameter region is defined by ${(\alpha_1, \alpha_2) \mid 0 \leq \alpha_2 < \alpha_1 < 1}$, as the uniqueness theorem applies only to non-degenerate solutions.

\medskip
In the Weyl-Papapetrou coordinates $(\rho,z)$, the rod structure, for both solutions,  our transformed soution~(\ref{eq:neutral-sol-cmetric-2}) and  Pomeransky-Sen'kov metric~(\ref{eq:metricPS}), can be sumarized as follows:

\medskip
 \begin{center}
 \begin{tabular}{|c|c|} \hline
Solutions & Rods  \\ \hline
  Our transformed solution& $(-\infty,-\nu \ell^2),  \ (-\nu \ell^2,\nu \ell^2), \ (\nu \ell^2,\ell^2), \ (\ell^2,\infty) $ \\ \hline
  Pomeransky-Sen'kov solution& $\left(-\infty, - \frac{\alpha_1-\alpha_2}{1-\alpha_1\alpha_2} k^2\right),  \  
  \left(- \frac{\alpha_1-\alpha_2}{1-\alpha_1\alpha_2} k^2, \frac{\alpha_1-\alpha_2}{1-\alpha_1\alpha_2} k^2\right), \  \left(\frac{\alpha_1-\alpha_2}{1-\alpha_1\alpha_2} k^2,k^2\right), \ (k^2,\infty) $ \\ \hline
 \end{tabular}
 \end{center}

\medskip
Then, two metrics have the rods of the same length  if
\begin{align}
\ell=k,\quad \nu = \frac{\alpha_1-\alpha_2}{1-\alpha_1\alpha_2}. \label{eq-ellk}
\end{align}
Moreover, as the third condition,  we compare the mass~(\ref{eq:mass}) for our transformed solution with the mass ~(\ref{eq:mass-PS}) for the Pomeransky-Sen'kov solution.
Using Eqs.~(\ref{eq:br-exp-b}), (\ref{eq:br-exp-beta}) and  (\ref{eq:a-exp-2}),  we can express the mass~(\ref{eq:mass}) as a simpler form,
\begin{align}\label{eq:mass2}
M = \frac{3 \ell^2 \pi  (\gamma +\nu ) \left(\nu(\gamma  -\nu)+
    \sqrt{2\nu(1-\nu^2 )  (\gamma -\nu )}\right)}{2 (\gamma
   -\nu ) \left(2-\nu^2-\gamma \nu\right)}.
\end{align}
Equating Eq.~(\ref{eq:mass2}) with Eq.~(\ref{eq:mass-PS}), along with Eq.~(\ref{eq-ellk}),
one can express $(\alpha_1,\alpha_2)$ as functions of $(\nu,\gamma)$
\begin{align}\label{eq:seta1a2}
\alpha_1 = U_+-\sqrt{U_+^2-1},\quad \alpha_2 =  U_- -\sqrt{U_-^2-1},
\end{align}
where
\begin{align}
U_\pm = \frac{\nu  \left(-\gamma ^2 (1\mp \nu )^2 + 2 \gamma  \left(\nu ^3\pm 2 \nu ^2+\nu \mp 4\right)
+\nu  \left(3 \nu ^3\pm 2 \nu ^2+3 \nu \pm 8\right)\right)
+2\sqrt{2} (\gamma +\nu )    \sqrt{\nu  \left(1-\nu ^2\right)^3 (\gamma -\nu )}}{\nu  (\gamma  (1\mp\nu )-\nu  (3\pm\nu ))^2}.
\end{align}
Note that the $\pm$ symbol in $U_{\pm}$ are not related to the $(\pm\pm)$-branches used in the previous section.

\begin{figure}
\includegraphics[width=6cm]{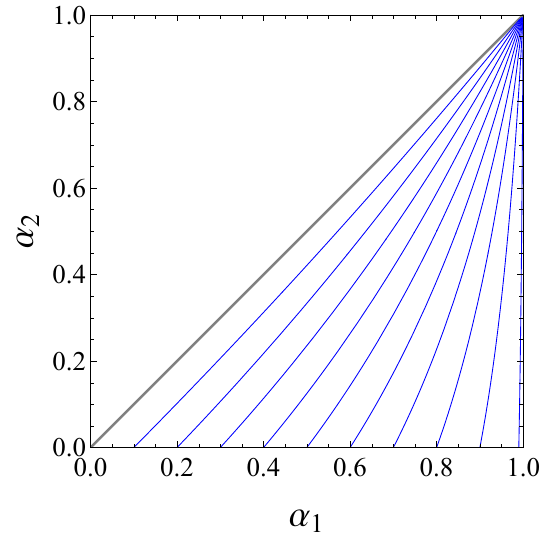}
\caption{$\nu={\rm const}.$ curves (blue curves) in the $(\alpha_1,\alpha_2)$ plane mapped by Eq.~(\ref{eq:seta1a2}). Each curves correspond to $\nu=1/10,2/10,\dots,9/10$ and $99/100$. $\gamma$ runs 
from $\nu(3-\nu)/(1+\nu)$ to $\nu$, from the bottom to the top. The $\nu={\rm const.}$ curve approaches the extremal limit $\alpha_1=\alpha_2$ ( gray line) at $\nu \to 0$.  \label{fig:map}}
\end{figure}

\medskip
It should be noted that  
Eq.~(\ref{eq:seta1a2}) does not provide a one-to-one mapping from the range of $\{ (\nu,\gamma) \mid 0<\nu <\gamma<1\}$ to the range of $\{ (\alpha_1,\alpha_2) \mid 0 \leq \alpha_2 < \alpha_1<1\} $,  because as shown in~Fig.\ref{fig:map}
the parameter subregion of the $(++)$-branch for our solution: $\left\{(\nu,\gamma) \mid \nu < \gamma \leq \nu(3-\nu)/(1+\nu),\ 0< \nu <1 \right\}$ has a one-to-one correspondence to the entire parameter region of the Pomeransky-Sen'kov solution: $\{ (\alpha_1,\alpha_2) \mid 0 \leq \alpha_2 < \alpha_1<1\}$. 
Moreover, using Eqs.~(\ref{eq:br-exp-b}), (\ref{eq:br-exp-beta}), (\ref{eq:a-exp-2}) and (\ref{eq:seta1a2}), one can easily show that the two angular momenta $|J_\psi|$ and $|J_\phi|$ for our solution, given by Eqs.~(\ref{eq:Jpsi-trans}) and (\ref{eq:Jphi-trans}) exactly coincide with those for the Pomeransky-Sen'kov solution, given by Eqs.~(\ref{eq:Jphi-PS}) and (\ref{eq:Jpsi-PS}). 
Therefore, it can be expected from the uniqueness theorem~\cite{Hollands:2007aj, Morisawa:2007di} that the two metrics, the $(++)$-branch metric for our solution and the Pomeransky-Sen'kov solution, are isometric, as long as the metric for the $(++)$-branch is regular and free from closed timelike curves on and outside the horizon.

\medskip
Actually, one can directly show that the $(++)$-branch metric~(\ref{eq:neutral-sol-cmetric-2}) coincides with the Pomeransky-Sen'kov metric~(\ref{eq:metricPS})
using Eqs.~(\ref{eq:br-exp-b}), (\ref{eq:br-exp-beta}), (\ref{eq:a-exp-2}) and (\ref{eq:seta1a2}) as follows.
As given in the Appendix~\ref{sec:PS}, the metric of the Pomeransky-Sen'kov solution is written in the C-metric coordinates $(x,y)$ as our transformed metric, but it should be noted that the choice of the C-metric coordinates $(x,y)$ is not unique. 
For the latter, the function $G(u)$ in Eq.~(\ref{eq:G}) can be written as a cubic function with respect to $u$, whereas for the former, the corresponding function~(\ref{eq:GPS}) can be expressed in a quartic function as
\begin{align}
G_4(u) = (1-u^2)(1+\alpha_1 u)(1+\alpha_2 u),
\end{align}
where the outer and inner horizons correspond to  $y=-1/\alpha_1$ and $y=-1/\alpha_2$.
These two metrics apparently differ, but are related by the coordinate transformation
 $(x,y)\to (x_{\rm new},y_{\rm new})$, referred to as  the M\"{o}bius transformation~\cite{Chen:2011jb}, 
which is defined by
\begin{align}
 x = \frac{x_{\rm new}+\alpha_2}{1+\alpha_2 x_{\rm new}},\quad y= \frac{y_{\rm new}+\alpha_2}{1+\alpha_2 y_{\rm new}},  \quad \nu = \frac{\alpha_1-\alpha_2}{1-\alpha_1\alpha_2},
 \label{eq:moebius-t}
\end{align}
which transforms the C-metric written in the cubic function to the C-metric written in the quartic function. 
\begin{align}
 \frac{dx^2}{G(x)} -  \frac{dy^2}{G(y)} = (1-\alpha_1 \alpha_2)\left[ \frac{dx_{\rm new}^2}{G_4(x_{\rm new})}-\frac{dy_{\rm new}^2}{G_4(y_{\rm new})}\right].
\end{align}
Thus, in the new coordinates $(x_{\rm new},y_{\rm new})$, by employing Eqs.~(\ref{eq:br-exp-b}), (\ref{eq:br-exp-beta}), (\ref{eq:a-exp-2}) and (\ref{eq:seta1a2}), it is evident that the $(++)$-branch for our metric~(\ref{eq:neutral-sol-cmetric-2}), derived via the Ehlers transformation, coincides with the Pomeransky-Sen'kov metric~(\ref{eq:metricPS}).
It should be noted that the same rationale applies to the other $(+-)$, $(-+)$ and $(--)$-branches as well.
\medskip

\section{Summary and Discussion}\label{sec:sum}

In this paper, by employing the sigma model approach, we have obtained an exact solution for a doubly rotating black ring to the five-dimensional vacuum Einstein equation. 
Giusto and Saxena have previously shown that the $SO(2,1)$ subgroup of the $SL(3,{\mathbb R})$ symmetry can produce a one-parameter family of stationary solutions with angular momenta. 
However, unlike black holes, merely applying this technique to the static black ring or the Emparan-Reall black ring does not resulted in a regular singly rotating black ring or a regular doubly rotating black ring, due to the emergence of a Dirac-Misner string singularity.
To tackle this problem, we have utilized a vacuum solution of a rotating black ring, which already includes a Dirac-Misner string singularity, as the seed solution for the Ehlers transformation.
The fundamental idea is similar to the construction of a charged black ring preformed in our previous work, where as a vacuum seed for the electric Harrison transformation, we chose the black ring solution with a Dirac-Misner string singularity. 
By appropriately adjusting the solution's parameters after the transformation, the resultant solution has proven to be regular, showing no signs of curvature singularities, conical singularities, orbifold singularities, Dirac-Misner string singularities, or closed timelike curves, both on and outside the horizon. 
As expected from the uniqueness theorem for a stationary and bi-axisymmetric black ring~\cite{Hollands:2007aj},  this solution exactly coincides with the Pomeransky-Sen'kov solution.

\medskip

According to the uniqueness theorem for charged black rings in the five-dimensional supergravity~\cite{Tomizawa:2009ua,Tomizawa:2009tb}, 
under the assumptions of the trivial topology (${\Bbb R} \times \{ {\Bbb R}^4 \setminus  D^2\times S^2 \} $) 
of the domain of outer communication, an asymptotically flat, stationary and bi-axisymmetric charged rotating black ring with non-degenerate connected event horizon is characterized uniquely by the mass, electric charge, two independent angular momenta, dipole charge, and additional information on the rod structure such as the ratio of the $S^2$ radius to the $S^1$ radius. 
Elvang {\it et al.}~\cite{Elvang:2004xi} constructed an example of such a dipole ring, which is originating from a seven-parameter family of non-supersymmetric black ring solutions, furthermore, we gave another example, which was constructed by the electric Harrison transformation. 
These both solutions carry the five physical quantities, but three only are independent among them.    
As conjectured by the authors in Ref.~\cite{Elvang:2004xi}, 
it is natural that a more general non-Bogomol’nyi-Prasad-Sommerfield (BPS) black ring solution exists, characterized by its mass, two independent angular momenta, electric charge, and a dipole charge, which are independent of the other asymptotic conserved charges, which finally may have been constructed in Ref.~\cite{Feldman:2014wxa}. 
However, this solution has its considerably complex and formal expression, at this point, we do not know whether this solution truly describes a regular black ring solution with the five independent parameters, including the regularity, the horizon topology, asymptotic flatness, as well as physical properties. 
Therefore, to investigate them, we need a more compact form of the solution, such as the C-metric form. 
We anticipate that applying our method to the five-dimensional minimal supergravity may enable us to derive such an exact solution.

\section*{Acknowledgement}
R.S. was supported by JSPS KAKENHI Grant Number JP24K07028.
S.T. was supported by JSPS KAKENHI Grant Number 21K03560.

\appendix

\section{Coefficients $k^\pm_1,\dots,k^\pm_9$}
\begin{align}
\begin{split}
&k_1^\pm= \frac{c_1 c_3 (\nu -1) (\gamma +\nu ) \left(2 (\gamma -1) c_1 \nu  (\nu +1)
\pm 2  (1-\gamma ) \nu  \left(\nu ^2-1\right)
\pm d_1 (1-\nu )\pm2 d_2 \nu \right)}{8 (\gamma -1)},\\
&   k_2^\pm= -\frac{(1\pm a) (1-\gamma ) d_1 (1-\nu )^3 (\gamma +\nu )}{4 (\nu +1)},\\
& k_3^\pm=   \frac{c_3 d_1 (1-\nu^2)  (\gamma +\nu ) \left((1-\gamma)(1-\nu)\pm4(\gamma-\nu)\nu b \right)}{8 (\gamma -1) \nu },\\
& k^\pm_4= \frac{c_3 d_1 (1-\nu ) (\gamma +\nu )}{8 \nu   },\\
&k^\pm_5= \frac{(1\pm a) (\gamma -1) c_3 (\nu -1)^3 (\gamma +\nu )}{4 (\nu +1)},\\
&k_6^\pm=   -\frac{\nu}{4}  \biggr[2   c_3^2 (\nu -1)^2 (\nu -\gamma )
-2 (1\pm a) (1-\gamma ) c_3 (\nu -1)^2 (2  (1-\gamma ) \nu\pm b (\nu +1) (\gamma -\nu ) )\\
&\quad   -(1\mp a) (1-\gamma ) d_1 (1-\nu ) \left(3   \gamma  \nu +\gamma -\nu ^2-3 \nu \pm 2 b (\nu +1) (\gamma -\nu )\right)+c_3 d_1 (\nu -1) (3\nu-\gamma \mp 2a(1-\gamma) )\biggr],\\
&   k_7^\pm= \frac{c_3 (1-\nu )^3}{8 (\nu +1)} \biggr[4 a (1-\gamma )  c_1 (\nu +1) \nu 
\pm 2 a (\gamma -1) (\nu -1) \nu  (\gamma +\nu )\pm 4 (\gamma -1) c_2 \nu \\ 
&\quad
   +2   (\gamma -1) (\nu -1) \nu  (\gamma -\nu -2)-d_1 (\gamma +\nu )\biggr],\\
&   k_8^\pm=
   \frac{d_1 (\nu -1)}{2 (\nu +1)} \biggr[ 2 a b (\gamma -1) (\nu +1) (\gamma -\nu )\pm (1-\gamma ) \left(a   (\nu -1) (\gamma +\nu )+2 c_2\right)\nonum
   &\quad +b^2 (\nu +1) \left(\nu ^2-\gamma ^2\right)+(1-\gamma
   ) (\nu +1) (\gamma -3 \nu ) \biggr]\\
&k^\pm_9 = \frac{(\gamma +\nu ) \left(1-a^2+\gamma  (b-a)^2-\nu  \left(1-2 a b+b^2\right)\right)}{8
   (1\mp a) (1-\gamma )}.
 \end{split}
\end{align}

\section{Pomeransky-Sen'kov's doubly-spinning black rings}\label{sec:PS}
Here we recapitulate the metric for the doubly-spinning black rings~\cite{Pomeransky:2006bd}\footnote{We have exchanged $\psi$ and $\phi$ so that $\psi$ generates the $S^1$-rotation.}
\begin{align}\label{eq:metricPS}
ds^2 &= - \frac{H_{\rm PS}(y,x)}{H_{\rm PS}(x,y)}(dt + \Omega_{\rm PS})^2 +\fr{H_{\rm PS}(y,x)}
\left[F_{\rm PS}(x,y)d\psi^2-2 J_{\rm PS}(x,y)d\psi d\phi-F_{\rm PS}(y,x)d\phi^2\right]\nonum
&+ \frac{2k^2H_{\rm PS}(x,y)}{(x-y)^2(1-\alpha_1 \alpha_2)^2}\left(\frac{dx^2}{G_4(x)}-\frac{dy^2}{G_4(y)}\right),
\end{align}
where
\begin{align}
\begin{split}
G_4(u)&=(1-u^2)(1+\alpha_1 u)(1+\alpha_2 u), \label{eq:GPS}\\
H_{\rm PS}(x,y)&=1-\alpha _1^2 \alpha _2^2+ 2 \left(1-x^2\right) y \alpha _1 \alpha _2 \left(\alpha _1+\alpha _2\right)
+\left(\alpha _1+\alpha _2\right){}^2   + 2 x \left(\alpha _1+\alpha  _2\right) \left(1-y^2 \alpha _1^2 \alpha _2^2\right)\\
&\quad+x^2 y^2 \alpha _1 \alpha _2   \left(1-\alpha _1^2 \alpha _2^2-\left(\alpha _1+\alpha _2\right){}^2\right),\\
 F_{\rm PS}(x,y)&=\frac{2 k^2}{(x-y)^2 \left(1-\alpha _1
   \alpha _2\right){}^2}\biggr[ G(y) \biggr\{ 2 \left(\alpha _1+\alpha _2\right){}^2+x \left(1+\alpha
   _1^2\right) \left(\alpha _1+\alpha _2\right) \left(1+\alpha _2^2\right)\\
&\quad \quad  +x^3   \left(\alpha _1+\alpha _2\right) \left(1-3 \alpha _1^2 \alpha _2^2+2 \alpha _1^3
   \alpha _2^3-\left(\alpha _1+\alpha _2\right){}^2\right)-x^4 \alpha _1 \alpha _2
   \left(1-\alpha _1 \alpha _2\right) \left(-1+\alpha _1^2 \alpha _2^2+\left(\alpha
   _1+\alpha _2\right){}^2\right)\\
&\quad \quad   +x^2 \left(1+\alpha _1 \alpha _2\right)
   \left(-\left(\alpha _1+\alpha _2\right){}^2+\left(-1+\alpha _1 \alpha
   _2\right){}^2\right)\biggr\}\\
&\quad   +\left(1-y^2\right) G(x) \left(\left(1+\alpha _1 \alpha
   _2\right) \left(-\left(\alpha _1+\alpha _2\right){}^2+\left(-1+\alpha _1 \alpha
   _2\right){}^2\right)-y \left(\alpha _1+\alpha _2\right) \left(-1+\alpha _2^2+\alpha
   _1^2 \left(1+3 \alpha _2^2\right)\right)\right)\biggr],\\
J_{\rm PS}(x,y)&=\frac{2 k^2 \left(1-x^2\right) \left(1-y^2\right) \left(\alpha
   _1+\alpha _2\right)\sqrt{\alpha _1 \alpha _2} }{(x-y) \left(1-\alpha _1 \alpha _2\right){}^2}
    \biggr[1+\left(\alpha _1+\alpha _2\right){}^2-\alpha _1^2 \alpha _2^2+2 \alpha _1 \alpha _2   \left(\alpha _1+\alpha _2\right)(x+y)\\
&\quad     + \alpha _1
   \alpha _2 \left((\alpha _1 +\alpha _2)^2+\alpha _1^2 \alpha
   _2^2-1\right) xy\biggr],\\
\Omega_{\rm PS}(x,y) &= -\frac{2 k  \left(\alpha _1+\alpha
   _2\right) \sqrt{\alpha_1\alpha_2\left(1-\alpha _1^2\right) \left(1-\alpha _2^2\right)}}{H(y,x)} \biggr[ \left(1-x^2\right) yd\phi \\
&\quad    -\frac{ (1+y)  \left(1+\alpha _1+\alpha _2-\alpha _1 \alpha _2+2 x
   (1-y) \alpha _1 \alpha _2-x^2 y \alpha _1 \alpha _2 \left(-1+\alpha _1+\alpha
   _2+\alpha _1 \alpha _2\right)\right)}{\left(1-\alpha _1\right) \left(1-\alpha
   _2\right)}d\psi\biggr].
\end{split}
\end{align}
For the convenience, we have replaced the original parameters $(\lambda,\nu)$ in Ref.~\cite{Pomeransky:2006bd} with the following $(\alpha_1,\alpha_2)$
\begin{align}
\lambda = \alpha_1+\alpha_2,\quad \nu = \alpha_1 \alpha_2,\quad 0\leq \alpha_2\leq \alpha_1<1. 
\end{align}
On the horizon at $y = -1/\alpha_1$, we obtain the horizon velocity
\begin{align}
 \omega^{\rm PS}_\psi = \frac{\sqrt{(1-\alpha_1^2)(1-\alpha_2^2)}}{2 k(1+\alpha_1)(1+\alpha_2)},\quad
 \omega^{\rm PS}_\phi = \frac{(1-\alpha_1)(1+\alpha_1^2)(1-\alpha_2)\alpha_2}{2k\sqrt{\alpha_1\alpha_2}(\alpha_1+\alpha_2)\sqrt{(1-\alpha_1^2)(1-\alpha_2^2)}}.
\end{align}
and the surface gravity
\begin{align}
  \kappa_{\rm PS} =\frac{ (1-\alpha_1)(\alpha_1-\alpha_2)(1-\alpha_1\alpha_2)}{4 k \alpha_1 (1+\alpha_2)(\alpha_1+\alpha_2)}.
\end{align}
The asymptotic charges are given by
\begin{align}
M_{\rm PS} &=\frac{3 k^2 \pi  \left(\alpha _1+\alpha _2\right)}{G_5\left(1-\alpha _1\right)
   \left(1-\alpha _2\right)}, \label{eq:mass-PS}\\
J^{\rm PS}_\phi &=  \frac{4 k^3 \pi  \sqrt{\alpha _1 \alpha _2} \left(\alpha
   _1+\alpha _2\right) \sqrt{\left(1-\alpha _1^2\right) \left(1-\alpha   _2^2\right)}}{G_5\left(1-\alpha _1\right) \left(1-\alpha _2\right) \left(1-\alpha _1   \alpha _2\right){}^2},\label{eq:Jphi-PS}\\
J^{\rm PS}_\psi &=   \frac{2 k^3 \pi  \left(\alpha _1+\alpha _2\right)
   \sqrt{\left(1-\alpha _1^2\right) \left(1-\alpha _2^2\right)} \left(1+\alpha
   _2+\alpha _1^2 \alpha _2 \left(1+\alpha _2\right)+\alpha _1 \left(1-6 \alpha _2+\alpha
   _2^2\right)\right)}{G_5\left(1-\alpha _1\right){}^2 \left(1-\alpha _2\right){}^2
   \left(1-\alpha _1 \alpha _2\right){}^2}.\label{eq:Jpsi-PS}
\end{align}

\section{Coefficients $f_i$ and $g_i$}\label{sec:fi-gi}
The explicit forms of the coefficients $f_i$ and $g_i$ are shown below. 

\subsection{$f_1,\dots,f_{19}$}
$f_i$ are quadratic functions of $\beta$. $f_i$ for $16\leq i\leq 19$ does not appear in the metric functions but does only in the expressions of $g_i$.
\begin{align}
 f_1& = 1+2( a- b) \beta +\frac{\beta ^2}{\gamma^2 -\nu^2 } \biggr[(a-b)^2 \gamma ^2-(1-a)^2 \gamma  (1+2 b+(3+2 b) \nu )\nonum
&\quad\quad+\nu  \left((1-a)^2 (3+2 b)+\left(1-2 a+2 \left(1-a+a^2\right) b-b^2\right) \nu \right)\biggr],\\
 f_2 &= 1+2(1- a b) \beta +\frac{\beta ^2}{\gamma^2 -\nu^2 } \biggr[ \left((2-a) a-2 \left(1-a+a^2\right) b-(1- 2 a) b^2\right) \gamma ^2\nonum
&\quad \quad +(1-a)^2 \gamma  (1+2 b+(3+2 b) \nu   )-\nu  \left((1-a)^2 (3+2 b)+(1-b) (1+(1-2 a) b) \nu \right)\biggr],\\
 f_3 &= 1+2( a- b) \beta +\frac{\beta ^2}{\gamma^2 -\nu^2 } \biggr[ (a-b)^2 \gamma ^2-2 (1-a)^2 (1+a)+(1-a)^2 (1-2 a+2 b) \nu \nonum
&\quad\quad+ \left(1-2 a+2 \left(1-a+a^2\right)   b-b^2\right) \nu ^2+(1-a)^2 \gamma  (1+2 a-2 b-(1-2 a+2 b) \nu )\biggr],\\
 f_4 &= 1- \frac{2 \beta}{\gamma -\nu }  \left(a^2-1-a (a-b) \gamma +\nu(1 -a b) \right)+\frac{\beta ^2}{\gamma^2 -\nu^2 } \biggr[ 2 (1-a)^2 (1+a)-(1-2 a) (a-b)^2 \gamma   ^2\nonum
&\quad \quad -(1-a) (1-5 a+4 b) \nu -\left(1-2 a b-b^2+2 a b^2\right) \nu ^2-(1-a) \gamma  \left(1-a-4 a^2+4 a b-(1 -5 a  +4 b) \nu   \right)\biggr],\\
 f_5& = 1   - \frac{2 \beta }{c_2} \left(2 a   (-1+\gamma ) \nu +a b^2 (\gamma -\nu ) (1+\nu )+b \left(-1+a^2-a^2 \gamma -\left(2+a^2\right) (-1+\gamma ) \nu +\nu ^2\right)\right)\nonum
& \quad +\frac{\beta ^2}{(\gamma^2 -\nu^2 )  c_2} \biggr[ b \gamma  \left(2 (1-a)^2 (1+a)+(1-a) \left(-1+a+4 a^2-4 a b\right) \gamma +(-1+2 a) (-a+b)^2 \gamma ^2\right)\nonum
& \quad +\nu \biggr\{   \left(2 \left(-2+a+6 a^2-4 a^3\right)+\left(4-20 a+17 a^2-6   a^3\right) b+4 \left(2-3 a+2 a^2\right) b^2+(1-2 a) b^3\right) \gamma ^2\nonum
& \quad \quad 
+(-a+b)^2 (-4+2 a-b+2 a b) \gamma ^3-2   (1-a)^2 (1+a) (2+b)+2 (1-a)^2 \left(4+5 a+(-1+3 a) b-2 b^2\right) \gamma\biggr\} \nonum
& \quad    +  \nu ^2 \biggr\{  (1-a)   \left(2 \left(2-3 a+a^2\right)+\left(3-9 a+2 a^2\right) b +4 b^2\right)\nonum
& \quad \quad +\left(4 (a-2) (1-a)^2+\left(-10+26 a-21 a^2+4 a^3\right) b-2   \left(4-7 a+2 a^2\right) b^2+(1-2 a) b^3\right) \gamma\nonum
& \quad \quad +\left(2 (2-a) (1-a)^2+\left(5-14 a+10 a^2-2 a^3\right) b+\left(4-6 a+4 a^2\right)   b^2+(1-2 a) b^3\right) \gamma ^2\biggr\}\nonum
& \quad   - \left(2 a-\left(6-6 a+5 a^2\right) b+4 a b^2+(1-2 a) b^3\right) (1-\gamma ) \nu ^3+(1-b) b (1+(1-2 a) b) \nu ^4 \biggr],
\end{align}
\begin{align}
 f_6&= 1-\frac{2 \beta}{(\gamma -\nu ) (1+\nu ) c_1}  \biggr[ a^2-1+\left(2-2 a+a^2+b-2 a b\right) \nu
 -(1-b) (1-2 a+b)   \nu ^2-(1-b) b \nu ^3\nonum
 &\quad\quad +(a-b)^2 \gamma ^2 (1+\nu )+\gamma  \left(1-2 a^2-b+2 a b-2   \left(1-a+a^2-2 a b+b^2\right) \nu +(1-b) (1-2 a+2 b) \nu   ^2\right)\biggr]\nonum
&\quad   +\frac{\beta ^2 }{(\gamma^2 -\nu^2 ) (1+\nu ) c_1}\biggr[
2 (1-a)^2   (1+a)-(1-a) \left(2-5 a-a^2+4 b\right) \nu \nonum
&\quad\quad+\left(-2-2 a+6 a^2-3 a^3+\left(3-4  a+3 a^2\right) b+(2-3 a) b^2\right) \nu ^2+\left(2-3 a+3 a^2 b-3 a b^2+b^3\right)
   \nu ^3+(1-b)^2 b \nu ^4\nonum
&\quad\quad +(a-b) \gamma ^2 \left(4 a^2-1-2a+(2-3 a) b+\left((2-a)^2-2 a b+b^2\right) \nu -\left(3 (1-a)^2+(2-a)b-b^2\right) \nu ^2\right)\nonum
&\quad\quad     -(a-b)^3 \gamma ^3 (1+\nu )+ (1-a) \gamma \left(a-2+5 a^2-4 a   b+\left(2-9 a-a^2+8 b\right) \nu\right) \nonum
&\quad\quad+\left(2+5 a-12 a^2+6   a^3-\left(8-12 a+7 a^2\right) b+3 a b^2-b^3\right) \gamma \nu ^2-\left(2-3 a+3 a^2 b-3 a   b^2+b^3\right) \gamma \nu ^3\biggr],\\
 f_7&=  1-\frac{2 \beta }{b (\gamma -\nu ) (1+\nu )} \biggr[(b-1) b \gamma +\left(2-2 a+b-b^2\right) (1-\gamma ) \nu
   +(1-b) b \nu ^2 \biggr]\nonum
&   +\frac{\beta ^2}{b   (\gamma^2 -\nu^2 ) (1+\nu )}
 \biggr[b \gamma    \left((1-a)^2+\left((2-a) a-2 b+b^2\right) \gamma \right)-(1-b)^2 b \nu ^3\nonum
 &\quad
 +\left(-(1-a)^2   (4+3 b)+4 (1-a) (1-2 a+(2-a) b) \gamma  +\left(4 (1-a) a-\left(4-6   a+a^2\right) b-2 b^2+b^3\right) \gamma ^2\right) \nu \nonum
 &\quad
   +\left(4 (-1+a)+(2-3 a) a b+2   b^2-b^3+(1-a) (4-b-3 a b) \gamma \right) \nu ^2\biggr],\\
 f_8&= 1+\frac{2 \beta}{(a-b) (\gamma -\nu )}\biggr[1-a^2+(a-b)^2 \gamma -\left(1-2 a b+b^2\right) \nu\biggr]\nonum
&   -\frac{\beta ^2}{(a-b)   \left(\gamma ^2-\nu ^2\right)}\biggr[
2(-1+ a+ a^2- a^3)-a \gamma( 1  + 2 a - 3 a^2) -(a^3-b^3) \gamma ^2-b^2 \gamma    \left(2+2 a^2+a (-4+3 \gamma )\right)\nonum
&\quad
+b \left(-2 (1-a)^2   (1+a)+\left(1+4 a-7 a^2+2 a^3\right) \gamma +3 a^2 \gamma ^2\right)\nonum
&\quad
- \nu (1-a) (\gamma -1) \left((a-5) a+\left(5-3 a+2   a^2\right) b+2 (1-a) b^2\right)  \nonum
&\quad
-\left(-2+a+\left(-1+6 a-2   a^2\right) b+\left(-2+a-2 a^2\right) b^2+b^3\right) \nu ^2\biggr],\\
 f_9&= 1+2 (a-b) \beta 
+\frac{\beta ^2}{\left(\gamma ^2-\nu ^2\right) c_1}\biggr[
-2 (1-a)^2 (1+a)-(a-b)^3 \gamma ^3+(1-a)^2  (4+a+2 b) \nu-(1-b) b (1-2 a+b) \nu ^3\nonum
& -\left(2-5 a+4 a^2+\left(1-2 a-a^2\right) b+a b^2\right) \nu   ^2
-(a-b) \gamma ^2 \left(1-a (2-b)-\left((1-a)^2-a   b+b^2\right) \nu \right)\nonum
&  -\gamma  \left((1-a)^2 (-2-3 a+2 b)+2 \left(2-3
   a+a^3\right) \nu +\left(-2+5 a-4 a^2+\left(-2+4 a+a^2\right) b-3 a b^2+b^3\right)
   \nu ^2\right)\biggr],\\
 f_{10}&=\frac{4 (1-a)  (1-\gamma ) d_1}{\gamma^2 -\nu^2 }\beta  (1+(a-b) \beta ),\\
 f_{11}&=1-\frac{2 \beta  (-1+a-a \gamma +b (-1+a+\gamma )+(-1+a b) \nu )}{\gamma +\nu   }\nonum
& +\frac{\beta ^2}{\gamma^2 -\nu^2 } \biggr[(a-b)^2 \gamma ^2
+ \nu  \left((1-a) \left(a-3+2  b^2\right)-(1-b) (1+b-2 a b) \nu \right)+(1-a) \gamma  \left(1+a-2  b^2+\left(3-a-2 b^2\right) \nu \right)\biggr],\\
f_{12}&=f_1+\frac{c_3 \left(f_1-f_3\right)}{(a+1) b (1-\gamma ) (\nu +1)},\\
f_{13}&=f_7+\frac{\left(b (\gamma-\nu^2) -(2+b) (1-\gamma ) \nu \right) f_{10}}{2 b (1+\nu ) d_1},
\end{align}
\begin{align}
 f_{14}&=1-\frac{2 \beta }{d_2} \biggr[ b \left(-1+a^2+\gamma -2 a^2 \gamma -(1-2 a) b \gamma   +(a-b)^2 \gamma ^2\right)\nonum
 &\quad+\left(-2+\left(2+a^2\right) b+(1-2 a) b^2-2   \left(-1+b+a^2 b-2 a b^2+b^3\right) \gamma +b (-a+b)^2 \gamma ^2\right) \nu
\nonum
&\quad   +\left(2-b-2 a b^2+b^3+\left(-2+b+(1+2 a) b^2-2 b^3\right) \gamma \right) \nu
   ^2+(-1+b) b^2 \nu ^3\biggr]\nonum
&   +\frac{\beta ^2}{\left(\gamma ^2-\nu ^2\right) d_2} \biggr[
b \gamma  \left(2 (1-a)^2   (1+a)+(1-a) \left(-2+a+5 a^2-4 a b\right) \gamma \right. \nonum
&\quad\quad\left. +\left(a \left(-1-2 a+4   a^2\right)+\left(1+4 a-7 a^2\right) b+(-2+3 a) b^2\right) \gamma ^2-(a-b)^3   \gamma ^3\right) \nonum
&\quad +\left(-2 (1-a)^2 (1+a) (2+b)+2 (1-a)^2 \left(5+6 a+4 a b-2   b^2\right) \gamma \right.\nonum
&\quad \quad\left. +\left(-6+4 a+16 a^2-12 a^3+\left(2-22 a+26 a^2-11 a^3\right)
   b+\left(9-16 a+11 a^2\right) b^2+(2-3 a) b^3\right) \gamma ^2\right.\nonum
&\quad\quad\left. -2 \left((3-2 a)
   a^2-3 a \left(2-2 a+a^2\right) b+\left(3-4 a+5 a^2\right) b^2-3 a b^3+b^4\right)
   \gamma ^3+b (-a+b)^3 \gamma ^4\right) \nu\nonum
&\quad  +\left((1-a) \left(2-6 a+4   a^2+\left(2-9 a+3 a^2\right) b+4 b^2\right)\right.\nonum
&\quad \quad
+\left(6 (1-a)^2 (-1+2   a)+\left(-12+37 a-34 a^2+8 a^3\right) b+\left(-13+20 a-5 a^2\right) b^2+(2-3 a)
   b^3\right) \gamma \nonum
&\quad \quad+\left(4-24 a+30 a^2-12 a^3+\left(18-39 a+32 a^2-7 a^3\right)
   b+2 \left(5-10 a+4 a^2\right) b^2\right) \gamma ^2\nonum
&\quad \quad\left.+\left(2 a \left(4-5 a+2   a^2\right)+\left(-8+13 a-10 a^2+2 a^3\right) b+\left(-3+4 a-3 a^2\right) b^2+(2+3
   a) b^3-2 b^4\right) \gamma ^3\right) \nu ^2\nonum
&\quad +\left(2 (a-2) a-\left(-10+14 a-10   a^2+a^3\right) b+\left(3-8 a+a^2\right) b^2+(-2+3 a) b^3\right.\nonum
&\quad \quad\left.
+2 \left(-1+4 a-2   a^2+\left(-8+14 a-10 a^2+a^3\right) b-\left(3-8 a+a^2\right) b^2-3 a
   b^3+b^4\right) \gamma \right.\nonum
&\quad \quad\left. +\left(2 (-1+a)^2-\left(-6+14 a-10 a^2+a^3\right)
   b+\left(5-8 a+a^2\right) b^2+(-2+3 a) b^3\right) \gamma ^2\right) \nu   ^3-(1-b)^2 b^2 \nu ^5\nonum
&\quad +\left(2-(2+a) b+\left(2-4 a+a^2\right) b^2+3 a b^3-b^4+\left(-2+(2+a)
   b-\left(1-4 a+a^2\right) b^2-(2+3 a) b^3+2 b^4\right) \gamma \right) \nu   ^4  \biggr],\\
f_{15}&=1-\frac{2 \beta }{(\gamma -\nu ) \left(c_3-(1-a b) (1-\gamma ) (1+\nu )\right)}\biggr[
1-a^2-\left(2+a^2+a \left(-2-2 b+b^2\right)\right) \nu \nonum
&+\left(1-2 (1-a) b-(1+a)   b^2+b^3\right) \nu ^2-(1-b) b^2 \nu ^3-(a-b) \gamma ^2 \left(a-b+b^2+\left(-2+a-b+b^2\right) \nu
   \right)\nonum
&+\gamma  \left(-1+2 a^2-2 a b+a b^2+2 \left((1-a)^2+b-2 a b+(1+a) b^2-b^3\right) \nu +\left(-1+2   (1-a) b+(2+a) b^2-2 b^3\right) \nu ^2\right)\biggr] \nonum
& +\frac{\beta ^2}{(\gamma^2 -\nu^2 ) \left(c_3-(1-a b) (1-\gamma )   (1+\nu )\right)}
 \biggr[ -2 (1-a)^2 (1+a) (1+b)\nonum
 &\quad-(1-a) \left(1+5 a-2 a^2+\left(-4+3 a-3 a^2\right) b-2 (1-a)   b^2\right) \nu \nonum
 &\quad +\left(4-6 a+3 a^2-\left(-2+8 a-4 a^2+a^3\right) b+\left(2-2 a+3 a^2\right) b^2-a b^3\right)
   \nu ^2 \nonum
 &\quad+\left(-1+(4-3 a) b+\left(-1+2 a+a^2\right) b^2-(2+a) b^3+b^4\right) \nu ^3+(-1+b)^2 b^2 \nu
   ^4 \nonum
 &\quad+(a-b)^2 \gamma ^3 \left(-1+2 a+(-2+a) b+b^2+\left(-3+2 a+(-2+a) b+b^2\right) \nu \right) \nonum
 &\quad+\gamma ^2
   \left(-1+2 a+4 a^2-6 a^3-\left(2+5 a-14 a^2+4 a^3\right) b+\left(4-10 a+3 a^2\right) b^2+a b^3\right.\nonum
   &\quad \quad \left. +\left(-2-4
   a+13 a^2-6 a^3+\left(8-22 a+18 a^2-5 a^3\right) b+\left(7-14 a+5 a^2\right) b^2+(2+a) b^3-b^4\right) \nu\right.\nonum
   &\quad \quad \left. 
   +\left(3 (-1+a)^2-(-2+a) (-1+a)^2 b+\left(1-4 a+2 a^2\right) b^2+2 b^3-b^4\right) \nu ^2\right) \nonum
 &\quad+\gamma 
   \left((-1+a) \left(-3+a+6 a^2+\left(-4-5 a+5 a^2\right) b-2 (-1+a) b^2\right)\right.\nonum
   &\quad \quad \left. +(-1+a) \left(-3-11 a+6
   a^2+\left(12-11 a+7 a^2\right) b-6 (-1+a) b^2\right) \nu\right.\nonum
   &\quad \quad \left.  -\left(7-12 a+6 a^2+\left(4-13 a+8 a^2-2 a^3\right)
   b+\left(5-6 a+5 a^2\right) b^2-(2+a) b^3+b^4\right) \nu ^2\right.\nonum
   &\quad \quad \left. +\left(1+(-4+3 a) b-\left(-1+2 a+a^2\right)
   b^2+(2+a) b^3-b^4\right) \nu ^3\right) \biggr], \\
   f_{16} &= 1-\frac{2 \beta  \left(-1+a^2+(a-b)^2 \gamma ^2+(1-2 a+b) \nu +(1-b) b \nu ^2-\gamma  \left(-1+2 a^2+b-2 a
   b+(1-2 a+b) \nu \right)\right)}{(\gamma +\nu ) c_1}\nonum
&   +\frac{\beta ^2}{\left(\gamma ^2-\nu ^2\right) c_1}\biggr[ 2 (1-a)^2 (1+a)-(a-b)^3 \gamma ^3-(1-a) \left(4+a^2-a (1+4 b)\right) \nu \nonum
&\quad+\left(2-3 a-\left(1-4 a+a^2\right) b+(-2+a) b^2\right) \nu   ^2+(1-b)^2 b \nu ^3\nonum
&\quad+(a-b) \gamma ^2 \left(-1-2 a+4 a^2+(2-3 a) b+\left((-1+a)^2+(2-3 a) b+b^2\right) \nu
   \right)\nonum
&\quad+\gamma  \left((1-a) \left(-2+a+5 a^2-4 a b\right)+2 (1-a) \left(2-a+a^2-4 a b+2 b^2\right) \nu\right)
\nonum
&\quad   -\left(2-3 a+(4-a) a b-(4-a) b^2+b^3\right)\gamma \nu ^2\biggr],  
   \end{align}
    \begin{align}
 f_{17}& =1+\frac{2 \beta  (1-a+(a-b) \gamma +(1-b) \nu )}{\gamma +\nu }\nonum
&+\frac{\beta ^2 \left((a-b)^2 \gamma ^2
+ \nu    \left((1-a) (1-3 a+2 b)-(1-b)^2 \nu \right)+(1-a) \gamma  (1+a-2 b-\nu +3 a \nu -2 b \nu   )\right)}{\gamma^2 -\nu^2},       \\
f_{18}& =1+\frac{2 \beta  \left(1-a^2+a (a-b) \gamma +(2-a) (a-b) (1-\gamma ) \nu -(1-a b) \nu ^2\right)}{(1-\nu )
   (\gamma +\nu )}\nonum
&   + \frac{\beta ^2}{(\gamma^2 -\nu^2 ) (1-\nu )} \biggr[ 2 (1-a)^2 (1+a)-(1-a) \left(3+2 a^2-a (1+4 b)\right) \nu +\left((a-2)   a+2 (3-2 a) a b-(3-2 a) b^2\right) \nu ^2\nonum
&\quad   +(1-b) (1+(1-2 a) b) \nu ^3+(a-b)^2 \gamma ^2 (-1+2 a+(-3+2 a)   \nu )\nonum
&\quad+(1-a) \gamma  \left(-1+a+4 a^2-4 a b+\left(2-2 a+4 a^2-8 a b+4 b^2\right) \nu +\left(-1+a-4 a b+4   b^2\right) \nu ^2\right) \biggr],   \\
f_{19}&=1+\frac{2 \beta  (1-a+(a-b) \gamma -(1-b)\nu  )}{\gamma -\nu }+\frac{\beta ^2 \left(1-a^2-2 b (1-a (1-\gamma ))+a^2 \gamma +b^2 \gamma -(1-b)^2 \nu \right)}{\gamma -\nu   }.
\end{align}


\subsection{$g_1,\dots,g_{11}$}

 \begin{align}
g_1 & =f_{13} f_{14}-\frac{c_1 c_3 \left(f_{13} f_{14}-f_{16} f_{19}\right)}{2 (a-b) (1-\gamma )^2 \nu }+\frac{(1-\nu )
   c_3 f_3 f_{10}}{2 (a-b) (1-\gamma ) (1+\nu ) d_1},
\label{eq:def-g1}\\
g_2&=D_0+\frac{c_3 \left(D_0-f_3 f_4\right)}{d_3}, \label{eq:def-g2}\\
g_3 &=f_{14}f_{18}-\frac{d_1 \left(f_{14} f_{18}-f_2 f_{17}\right)}{(1-\gamma ) (1+\nu ) \left(a   c_1-1+\nu \right)}-\frac{b \nu  (\gamma -\nu ) f_{10} f_{12}}{(1-\gamma ) (1-\nu^2 ) \left(a   c_1-1+\nu \right)},\\
g_4 &=D_0+\frac{1}{d_4}\biggr[-(1-\gamma )^2 (1-\nu )^2 \nu  d_1 \left(D_0-f_1^2\right)
-\frac{8 \nu  d_1 \left(d_1-(1+a) (1-\gamma )   (1+\nu ) c_1\right) \left(D_0-f_1 f_2\right)}{1+a}\nonum
&+2 (1-\gamma ) (1-\nu ) \nu  \left(-2-3 \nu +3 \nu   ^2\right) d_1 \left(D_0-f_2^2\right)-\frac{1}{2} (1-\nu ) (3-\nu ) d_2^2
   \left(D_0-f_{14}^2\right)\nonum
&-\frac{2 (\gamma -\nu ) (1-\nu )^3 c_3 \left(c_3+ b(1+a)  (1-\gamma ) (1+\nu   )\right) \left(D_0-f_3 f_4\right)}{1+a}+2 (1-\gamma ) (1-\nu )^2 \nu  c_2^2   \left(D_0-f_5^2\right)\nonum
&+\frac{1}{2} (\gamma -\nu ) (1+\nu ) \left(3-2 \nu +\nu ^2\right) c_1^2 c_3
   \left(D_0-f_6^2\right)
   +(1-\nu ) \nu ^2 \left(c_3-2 (1-\gamma ) \nu \right) d_1 \left(D_0-f_7^2\right)\nonum
   &+2
   (a-b)^2 (1-\gamma )^3 (1-\nu ) (\gamma -\nu ) \nu ^3 \left(D_0-f_8^2\right)+\frac{(\gamma -\nu ) (-1+\nu )^2
   \nu ^2 (\gamma +\nu ) c_3 f_{10}^2}{(1+\nu ) d_1}\biggr], \label{eq:def-g4}\\
g_{5} & =D_0-\frac{d_1 \left(D_0-f_1 f_2\right)}{a d_1+\left(1-a^2\right) (1-\gamma ) (1+\nu ) c_1},\\   
g_6 &=f_8^2+\frac{2 (1-\gamma )^2 (1-\nu )^2 \nu  d_1 \left(f_1^2-f_8^2\right)}{(1+\nu ) d_5}+\frac{(1-\gamma )^2   (1-\nu )^4 c_3 \left(f_3^2-f_8^2\right)}{(1+\nu ) d_5},\\
g_7 &= g_8 +\frac{c_3 f_{10}^2 (\nu -1)^2 \nu  \left(\gamma ^2-\nu ^2\right)}{d_1 d_6 (\nu +1)},\\
g_8 &=f_{14}^2-\frac{2 (1-\gamma ) (1-\nu )^2 \nu  d_1 \left(f_{14}^2-f_2^2\right)}{d_6}+\frac{(\gamma -\nu )
   (1-\nu )^2 (1+\nu ) c_1^2 c_3 \left(f_{14}^2-f_6^2\right)}{2 d_6 \nu },\\
g_9&=f_8 f_{18} +\frac{(2+\nu ) f_1 f_{10}}{(a-b) (1-\gamma ) (1-\nu^2 )},\\
g_{10}&=f_{14} f_{18}-\frac{b f_{10} f_{12} (\nu +1) (\gamma -\nu )}{2 d_2 (\nu -1)},\\
g_{11}&=f_{16}  f_{15}-\frac{(1-\gamma ) \left(4+\nu -\nu ^2\right) f_1 f_{10}}{(1+\nu ) \left(2 (a-b) (1-\gamma )^2   (1-\nu ) \nu -b (1+\nu ) d_1\right)}-\frac{(a-b) (1-\gamma )^2 (1-\nu )^2 \left(f_{18} f_8-f_{16} f_{15}\right)}{2 (a-b) (1-\gamma )^2 (1-\nu ) \nu -b (1+\nu ) d_1}.
\end{align}

\end{document}